\documentclass[aps,prb,showpacs,showkeys,twocolumn]{revtex4-1}
\usepackage[T2A,OT1]{fontenc}
\usepackage[cp1251]{inputenc}         
\usepackage[russian,ukrainian,english]{babel} 
\usepackage[dvips]{epsfig}            
\usepackage{latexsym}
\usepackage{amsfonts}
\usepackage{xcolor}[2007/01/21] 
\usepackage[dvipdfm,bookmarks=true,bookmarksnumbered,bookmarksopen]{hyperref} 
\hypersetup{pdfpagemode=FullScreen,colorlinks=true, citecolor=green,unicode,bookmarksopenlevel=3,
pdftoolbar=true,pdfauthor={Helen Gomonay},pdftitle={Spin transfer in AFM}}
\begin{document}

\title{Spin transfer and current-induced switching in antiferromagnets}

\author{Helen V. Gomonay}
\affiliation {
 National Technical University of Ukraine ``KPI''\\ ave Peremogy, 37, 03056, Kyiv,
Ukraine}

\author{Vadim M. Loktev}
\affiliation {Bogolyubov Institute for Theoretical Physics NAS of
Ukraine,\\ Metrologichna str. 14-b, 03143, Kyiv, Ukraine} 
\date{\today}

\begin{abstract}
Recent experiments show that spin-polarized current may influence
the state of  generally accessory element of spin-valves -- an
antiferromagnetic (AFM) layer, which is used for ``pinning''. Here
we study the dynamics of AFM component of the ``pinned''
ferromagnetic (FM) layer induced by simultaneous application of
the spin-polarized current and external magnetic field.

We find stability range of such a configuration of FM/AFM system
in which orientation of FM magnetization is parallel to AFM
vector. We calculate the field dependence of the critical current
for different orientations of the external magnetic field with
respect to the crystal axes of FM/AFM bilayer.  We show the
possibility of stable current-induced precession of AFM vector
around FM magnetization with the frequency that linearly depends
on the bias current. Furthermore, we estimate an optimal duration
of the current pulse required for switching between different
states of FM/AFM system and calculate the current and field
dependencies of switching time. The results obtained reveal the
difference between dynamics of ferro- and antiferromagnets
subjected to spin transfer torques.
\end{abstract}

\maketitle


\section{Introduction}
Spin transfer torque (STT) is the torque that is applied by
non-equilibrium spin-polarized conduction electrons onto a
magnetic layer \cite{Berger:1996,SLonczewski:1989,
  Slonczewski:1996}. This effect creates, for example, the ability to switch
  nanoscale magnetic devices at GHz frequencies and stimulate
  emission of microwaves by steady electric current \cite{Kaka:2005,
Slavin:2009}. Key elements of the spintronic devices, that enable
an information coding, control and manipulation by an electric
current, are two ferromagnetic (FM) layers. The ``pinned'' layer
acts as a polarizer for conduction electrons, while the state of a
``free'' layer may be altered by STT. However, recent experiments
\cite{Tsoi:2007, Urazhdin:2007, tang:122504, Dai08, Tsoi-2008}
give an indirect evidence that the spin-polarized current may also
influence the state of another, generally accessory, element -- an
antiferromagnetic (AFM) layer, which is used for ``pinning''.

The characteristic value of current density at which the switching
of AFM state takes place varies from 10$^5$~A/cm$^2$ (for an
insulating AFM and current-in-plane geometry \cite{Dai08}) to
10$^8$~A/cm$^2$ (for a metallic AFM and
current-perpendicular-to-plane geometry \cite{Tsoi:2007}) and, in
principle, can be smaller than the critical current density in the
similar giant magnetoresistive structures without an AFM layer
($\propto 10^7\div 10^9$~A/cm$^2$,
Refs.\onlinecite{Tsoi:1998,Grollier:2003,stiles-2008-320}).
Spin-polarized current also affects both the exchange bias and the
coercive field of ``free'' FM layer \cite{Basset08}. Combined
application of spin-polarized current and external magnetic field
gives rise to various switching scenarios depending on the
thickness, sequence and material of FM and AFM layers
\cite{Basset08}.
On the other hand, the physical mechanism and
details of such a nontrivial dynamics are still unclear.

Due to the efforts of several theoretical groups
\cite{tserkovnyak-2006, Nunez:2005, Haney:2007(2),xu:226602} the
concept of STT is extended to the systems with {\textit{i}})
different types of magnetic ordering, including nonuniform and
disordered FM (that, in principle, could be further extended to
the magnetic systems with noncollinear and, probably, AFM spin
ordering); {\textit{ii}}) different nature of the magnetic
ordering and interaction between the charge carriers and spins,
i.e., $sd$-exchange in the magnets with localized spins or
itinerant magnetism in the transition metals. It also became clear
that the STT phenomena could result from the atomic scale spin
dependent scattering (i.e., hopping of a conduction electron
between the sites with different directions of the magnetic
moments). In some particular cases an AFM can even work as a
polarizer for conduction electrons and exert spin torque on the
adjacent FM or AFM layer, as it was predicted in
Refs.\onlinecite{Nunez:2005, Haney:2007(2)}. However,  all of the
published calculations are based on assumption of quantum
coherence and so, are applicable to the perfect samples with ideal
interfaces.

In our previous paper \cite{gomo:2008E} we proposed the
phenomenological model that describes the current-induced
phenomena in AFMs on the same footing as in FM materials. It was
assumed that the total angular momentum is conserved during an
interaction of spin-polarized transport electrons with each of
magnetic sublattices.

In the present paper we apply this model for the description of
the precessional switching processes induced by simultaneous
application of the spin-polarized current and external magnetic
field to an AFM component of the ``pinned '' layer depicted in
Fig.~\ref{fig_geometry_new_2}(a). Our chief aim is to study the
different static and dynamic regimes of AFM layer and to find the
way to induce a stable precession of an AFM vector starting from a
certain configuration of FM/AFM bilayer. We also try to find
similar and different features in the current-induced dynamics of
FM/FM and FM/AFM bilayers. We anticipate our approach to be a
starting point for a more comprehensive analysis of the
multilayered magnetic systems in the presence of high-density
current. For example, joint behavior of the FM and AFM layers
could be analyzed, with account of the exchange bias coupling.

\section{Spin transfer torque in the multisublattice magnets}
According to Berger \cite{Berger:1996} and Slonczewski
\cite{SLonczewski:1989,
  Slonczewski:1996}, the physical mechanism of
STT in ferromagnets can be explained in the
  following way. When a free
electron transverses (or reflects from) an interface between the
nonmagnetic (NM) and FM layers, its spin state can be reversed
due to exchange interaction with the localized magnetic moments of
FM. This process results in rotation of the localized moments in a
way that ensures conservation of the total spin of the system.

Generalization of the Berger's and Slonczewski's ideas to the case
of multisublattice materials is not straightforward due to more
complex character of the magnetic ordering.  Particularly, in AFMs
the direction of the atomic magnetic moments varies on the length
scale of atomic distances leading to zero net magnetization if
averaged over few lattice constants.
However, just as in FMs, the spin-polarized electrons transfer
spin torques on each of atomic sites \cite{Nunez:2005, Haney:2007,
haney:2007(3),xu:226602}.

The magnetic structure of AFM may be described with the use of a
few macroscopic vectors $\mathbf{M}_j$ (in the simplest case
$j=1,2$) called the sublattice magnetizations (per unit volume)
that are formed due to  strong exchange coupling. So, it seems
reasonable to assume that while entering an AFM, the conduction
electron transfers spin angular momentum to any of the magnetic
sublattices (see Fig.~\ref{fig_geometry_new_2}b). Corresponding
STT $\mathbf{T}_{j}$ exerted by the $j$-th sublattice is then
presented in a standard form as follows:
\begin{equation}\label{STT_K}
  \mathbf{T}_{j}=\frac{\sigma_j
J}{M_{0j}}[\mathbf{M}_{j}\times[\mathbf{M}_{j}\times\mathbf{p}_{\rm
cur}]],
\end{equation}
 where $J$ is the current spin-polarized in
$\mathbf{p}_{\rm cur}$ direction, $\vert\mathbf{p}_{\rm
cur}\vert=1$, the constant $\sigma_j =\varepsilon \hbar
\gamma/(2M_{0j}Ve)$ is proportional to the efficiency
$\varepsilon$ of scattering processes, $V$ is the volume of AFM
region, $\hbar$ is the Plank constant, $e$ is the electron charge,
$\gamma$ is the modulus of the gyromagnetic ratio,  and
$M_{0j}=\vert \mathbf{M}_{j}\vert$ is the saturation magnetization
of $j$-th sublattice (the value of $M_{0j}$ is supposed to be
unchanged under external fields). Positive current ($J>0$)
corresponds to injection of electrons into AFM layer.

Then, the dynamics of  AFM can be described by a set of
Landau-Lifshitz-Gilbert equations for $\mathbf{M}_j$ vectors
supplemented with the Slonczewski term (\ref{STT_K}):
\begin{eqnarray}\label{Landau-Lifshitz}
  \dot{\mathbf{M}_j}&=&-\gamma[\mathbf{M}_j\times
  \mathbf{H}_j]+\frac{\alpha_G}{M_{0j}}[\mathbf{M}_j\times
\dot{\mathbf{M}_j}]\nonumber\\ &+&\frac{\sigma_j
J}{M_{0j}}[\mathbf{M}_{j}\times[\mathbf{M}_{j}\times\mathbf{p}_{\rm
cur}]],
\end{eqnarray}
where  $\mathbf{H}_j\equiv-\partial w/\partial \mathbf{M}_j$ is
the ``generalized force'' (an effective local field acting on the
magnetic moment of a sublattice) and $w$ is free energy (per unit
volume) of an AFM layer. For the sake of clarity we describe
relaxation of an AFM layer in the simplest form, with the use of a
single Gilbert damping parameter $\alpha_G$ equal for all magnetic
sublattices (although the relaxation mechanisms in AFM crystals
are very complicated and diverse \cite{Bar'yakhtar:1984E}.)

The last two terms in r.h.s. of Eq.~(\ref{Landau-Lifshitz}) are
responsible for dissipation processes in the AFM layer.  To
illustrate this fact we calculate the rate of free energy losses
in assumption that dissipation is small and in zero approximation
$\dot{\mathbf{M}_j}=-\gamma[\mathbf{M}_j\times
  \mathbf{H}_j]$. Thus,
\begin{eqnarray}\label{energy_losses_2}
  \frac{dw}{dt}&=&-\sum_j\left(\mathbf{H}_j\cdot\dot{\mathbf{M}_j}\right)\\
  &=&-\sum_j\left[\frac{\alpha_G}{\gamma
  M_{0j}}\dot{\mathbf{M}_j}^2-\frac{\sigma_j
J}{\gamma M_{0j}}\left(\mathbf{p}_{\rm
cur}\cdot[\mathbf{M}_j\times
\dot{\mathbf{M}_j}]\right)\right].\nonumber
\end{eqnarray}

In principle, Eqs.~(\ref{Landau-Lifshitz}),
(\ref{energy_losses_2}) could be used for description of different
complicated magnetic structures (compensated AFMs, weak FMs,
ferrimagnets). In the limiting case of the completely equivalent
sublattices ($\mathbf{M}_1=\mathbf{M}_2=\ldots$) the set of
equations (\ref{Landau-Lifshitz}) turns into a standard
Landau-Lifshitz-Gilbert-Slonczewski equation for FMs.

In the particular case of AFM with two magnetic sublattices it is
more suitable to rewrite Eqs.~(\ref{Landau-Lifshitz}) in terms of
macroscopic magnetization (FM vector)
$\mathbf{m}\equiv\mathbf{M}_1+\mathbf{M}_2$ and AFM order
parameter (AFM vector)
$\mathbf{l}\equiv\mathbf{M}_1-\mathbf{M}_2$:
\begin{eqnarray}
&&\dot{\mathbf{m}}=\gamma\left([\mathbf{H}_M\times
\mathbf{m}]+[\mathbf{H}_L\times
\mathbf{l}]\right)+\frac{\alpha_G}{2M_0}\left([\mathbf{m}\times\dot{\mathbf{m}}]+[\mathbf{l}\times\dot{\mathbf{l}}]\right)\nonumber\\
&&+\frac{\sigma
J}{2M_{0}}\left([\mathbf{m}\times[\mathbf{m}\times\mathbf{p}_{\rm
cur}]]+[\mathbf{l}\times[\mathbf{l}\times\mathbf{p}_{\rm
cur}]]\right)\label{Landau-Lifshitz_2}\\
&&\dot{\mathbf{l}}=\gamma\left([\mathbf{H}_M\times
\mathbf{l}]+[\mathbf{H}_L\times
\mathbf{m}]\right)+\frac{\alpha_G}{2M_0}\left([\mathbf{m}\times\dot{\mathbf{l}}]+[\mathbf{l}\times\dot{\mathbf{m}}]\right)\nonumber\\
&&+\frac{\sigma
J}{2M_{0}}\left([\mathbf{m}\times[\mathbf{l}\times\mathbf{p}_{\rm
cur}]]+[\mathbf{l}\times[\mathbf{m}\times\mathbf{p}_{\rm
cur}]]\right)\label{Landau-Lifshitz_2a}
\end{eqnarray}
Here  $\mathbf{H}_M=-\partial w/\partial \mathbf{m}$ is an
effective magnetic field within an AFM layer that includes an
external magnetic field, $\mathbf{H}_L=-\partial w/\partial
\mathbf{l}$ is a magnetic anisotropy field conjugated to an AFM
order parameter, and $|\mathbf{M}_1|=|\mathbf{M}_2|=M_0$.

Equations (\ref{Landau-Lifshitz_2}), (\ref{Landau-Lifshitz_2a})
describe the dynamics of FM and AFM vectors in the presence of
spin-polarized current and generalize the
Landau-Lifshitz-Gilbert-Slonczewski equation for the systems with
more than one magnetic sublattice. In what follows we base our
considerations on these equations.

\section{Model}
\begin{figure}[htbp]
 \includegraphics[width=1\columnwidth]{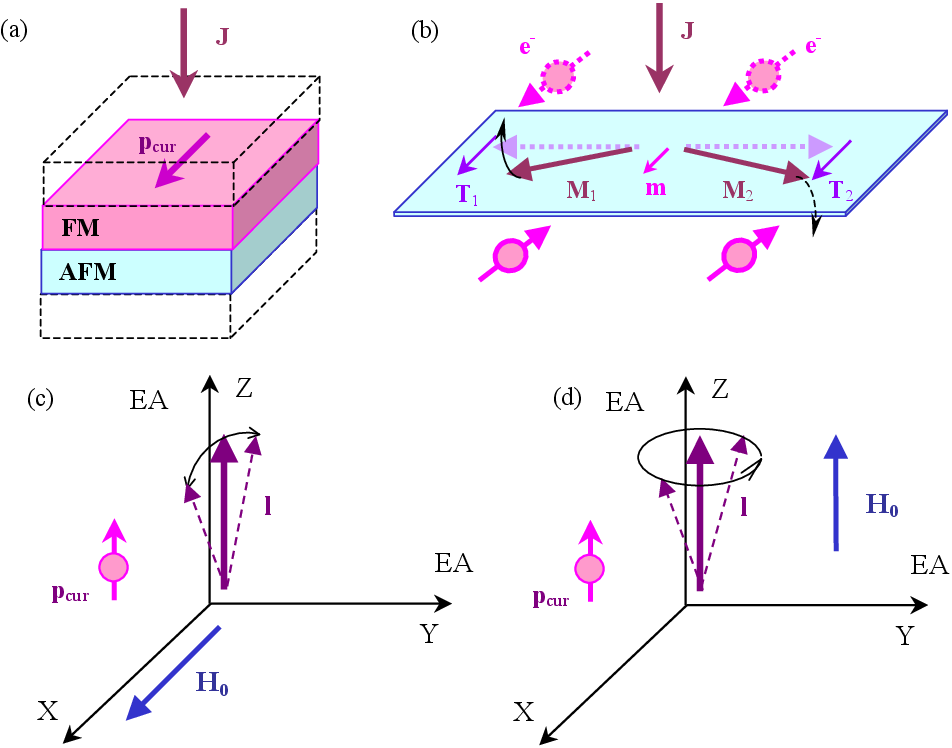}
  \caption{(Color online) \textbf{Effect of spin transfer torque within a pinned layer}.
  (a)
General structure of a pinned layer. Due to exchange coupling, an
AFM layer pins the direction of magnetization $\mathbf{p}_{\rm
cur}$ of the adjacent FM layer. In turns, the FM layer polarizes
the spin current flowing to AFM. (b) Transfer of spin torques
$\mathbf{T}_{1,2}$ from free electrons ($e^-$) to sublattice
magnetizations $\mathbf{M}_1$ and $\mathbf{M}_2$ (solid arrows --
before and dotted arrows -- after an interaction with the
conduction electrons). STT produces small FM moment $\mathbf{m}$
which, in turn, sets up a solid-like rotation of  $\mathbf{M}_1$,
$\mathbf{M}_2$ around $\mathbf{m}$ (arc arrows), according to
Eq.~(\ref{magnetization}). (c), (d) Sketch of
 eigen modes of an AFM vector for different orientations of the
external magnetic field. Easy axes (EA) are parallel to the film
plane.\label{fig_geometry_new_2}}
\end{figure}

Let us consider a pinned layer (Fig.~\ref{fig_geometry_new_2}a) of
a typical exchange-bias spin-valve that includes an AFM layer
whose thickness $d_{\rm AFM}$ is much smaller than the
characteristic scale of the magnetic inhomogeneity. On the other
hand, $d_{\rm AFM}$ is large enough to ensure an AFM ordering
within the layer. High current densities are achieved in a small
region (10$\div$100~nm in diameter), in which both FM and AFM
layers could be considered as a single domain. In the case of a
moderate pinning (i.e., when the magnetic anisotropy of FM is
comparable with the unidirectional anisotropy induced by exchange
bias) the FM works as a spin-polarizer whose state is not affected
by the precession of AFM vector in the adjacent layer. So,
magnetization of the FM layer is assumed to be fixed and is
described by the vector $\mathbf{p}_{\rm cur}$.

In our analysis we take into account the fact that AFMs (e.g.,
FeMn, IrMn) widely used in spintronic devices show strong exchange
coupling (corresponding exchange field $H_E\gg H_L$) between the
different magnetic sublattices that keeps magnetizations
$\mathbf{M}_1$ and $\mathbf{M}_2$ almost antiparallel even in the
presence of the external field $H_0\ll H_E$. In this case the
state of AFM is described by the only vector order parameter
$\mathbf{l}$, and far below the N\`{e}el temperature
$|\mathbf{l}|\approx 2M_0$. Spin-polarized current and/or external
magnetic field induce small tilt of the sublattice magnetizations
(Fig.~\ref{fig_geometry_new_2}b) formally described by the FM
vector $\mathbf{m}$. Vector $|\mathbf{m}|\ll |\mathbf{l}|$ is a
slave variable and can be expressed from
Eq.~(\ref{Landau-Lifshitz_2a}) as follows (see Appendix \ref{A}
for details):
\begin{equation}\label{magnetization}
\mathbf{m}=\frac{[\dot{\mathbf{l}}\times\mathbf{l}]}{2\gamma
M_0H_E}+\frac{1}{2M_0H_E}[\mathbf{l}\times[\mathbf{H}_0\times\mathbf{l}]],
\end{equation}
where $\mathbf{H}_0$ is the external magnetic field (which,
particularly, can be induced by Oersted field of a current).

Substitution of the expression (\ref{magnetization}) into
Eq.~(\ref{Landau-Lifshitz_2}) gives rise to a closed equation for
AFM vector:
\begin{eqnarray}\label{motion_equation}
&&[\ddot{\mathbf{l}}\times\mathbf{l}]=\gamma\left\{\left(2\dot{\mathbf{l}}(\mathbf{H}_0,\mathbf{l})-[\mathbf{l}\times
[\dot{\mathbf{H}}_0\times\mathbf{l}]]\right)\right.\nonumber\\
&&\left.-\gamma[\mathbf{H}_0\times
\mathbf{l}](\mathbf{H}_0,\mathbf{l})+2\gamma
M_0H_E[\mathbf{H}_L\times \mathbf{l}]\right.\nonumber\\ &&\left.
+\alpha_GH_E[\mathbf{l}\times\dot{\mathbf{l}}]+\sigma
JH_E[\mathbf{l}\times[\mathbf{l}\times\mathbf{p}_{\rm
cur}]]\right\}.
\end{eqnarray}
Equation (\ref{motion_equation}) describes a
solid-like motion of AFM vector in which $|\mathbf{l}|$ is almost
unchanged. Nevertheless, due to additional (compared to FM)
degrees of freedom, this equation differs from the standard
Landau-Lifshitz-Gilbert-Slonczewski equation for FM vector
$\mathbf{M}$ (see (\ref{Landau-Lifshitz}) for $j$=1) in order
(includes the 2-nd order derivative of the dynamic variable,
$\ddot{\mathbf{l}}$ instead of $\dot{\mathbf{M}}$).

Analysis of the last term in (\ref{motion_equation}) shows that
the current-induced contribution is proportional to the AFM order
parameter (vector $\mathbf{l}$). Thus, one can expect the
influence of spin-polarized current on the dynamics of AFM vector
to be at least as strong as in FM materials (other things being
equal). Moreover, the current-dependent (STT) term in the r.h.s.
of Eq.~(\ref{motion_equation}) contains large multiplier $H_E$.
This is manifestation of the so-called effect of exchange
enhancement when some interactions (e.g., gap values, spin-phonon
coupling) are more pronounced in AFMs than the analogous
interactions in FMs \cite{Bar:1968E}.

\section{Dynamics of AFM vector within the Lagrange approach}
An effective formalism for investigation of AFM dynamics is based
on the use of Lagrange formalism \cite{Bar-june:1979E}. Equation (\ref{motion_equation}) can be regarded as an
Euler-Lagrange equation of the second kind in the presence of
dissipative external forces (see Appendix \ref{B}).
 Corresponding Lagrange
function  has a form
\begin{widetext}
\begin{equation}
\label{Lagrangian} \mathcal{L}_{\rm AFM} =
\frac{1}{4\gamma^2M_0H_E }{\rm {\bf \dot {l}}}^2 -\frac{1}{2\gamma
M_0H_E}\left(\mathbf{H}_0\cdot[\mathbf{l}\times\dot{\mathbf{l}}]\right)+\frac{1}{4M_0H_E}[\mathbf{l}\times\mathbf{H}_0]^2-
w_{an} (\mathbf{l}).
\end{equation}
\end{widetext}
Here $w_{an}(\mathbf{l})$ is the energy of magnetic anisotropy
(per unit volume).

To take into account the effect of STT that can work both as a
source or drain of energy for an AFM layer, we deduce from
(\ref{energy_losses_2}) the dissipative Rayleigh function (see
Appendix \ref{B})
\begin{equation}
\label{Relay} \mathcal{R}_{\rm AFM} = \frac{\alpha_G}{4\gamma
M_0}\dot{\mathbf{l}}^2-\frac{\sigma J}{2\gamma
M_0}\left(\mathbf{p}_{\rm
cur}\cdot[\mathbf{l}\times\dot{\mathbf{l}}]\right)
\end{equation}
that describes the rate of the energy losses
\begin{equation}\label{energy_losses}
  \frac{dw}{dt}\equiv-\dot{\mathbf{l}}\cdot\left(\frac{\partial \mathcal{R}_{\rm AFM}}{\partial
  \dot{\mathbf{l}}}\right).
\end{equation}

Analysis of dissipative function (\ref{Relay}) shows that STT
phenomena in AFM have one general property which is not peculiar
to FM. While STT always changes the energy of FM layer, some types
of motions in AFM could be nondissipative even in the presence of
spin-polarized current. Linearly polarized oscillations of the
vector  $\mathbf{l}$, sketched  in Fig.~\ref{fig_geometry_new_2}c,
give an example of nondissipative mode (neglecting the internal
damping). And, vice versa, the most effective energy pumping
induced by the current takes place for any precessional, circular
polarized motion of AFM vector in the plane perpendicular to the
direction of current polarization $\mathbf{p}_{\rm cur}$ (see
Fig.~\ref{fig_geometry_new_2}d).

\section{Stability diagram}
To illustrate the peculiarities of non-dissipative and dissipative
current-induced dynamics, we analyze stability of the state with
parallel orientation of AFM and FM vectors, $\mathbf{p}_{\rm
cur}\|\mathbf{l}$, for two different configurations of the
external magnetic field $\mathbf{H}_0$, depicted schematically in
Figs.~\ref{fig_geometry_new_2}c,d. For the definiteness, an AFM
layer is supposed to have slightly tetragonal (almost cubic)
anisotropy induced, e.g., by the shape effects or/and interaction
with the neighboring layers (including possible influence of the
exchange bias). Two easy axes ($Z$ and $Y$) are parallel to the
film plane. In this case the magnetic anisotropy energy $w_{\rm
an}$ is modeled with the following expression
\begin{equation}\label{anisotropy}
w_{\rm an}=\frac{H_{\rm{an}\perp}}{M_0}l_X^2-\frac{H_{\rm{an}\|}}{8M_0^3}(l_X^4+l_Y^4+l_Z^4),
\end{equation}
where $H_{\rm{an}\|}$ is the intrinsic anisotropy field within the film plane  and the small out-of-plane
anisotropy field $H_{\rm{an}\perp}\ll H_{\rm{an}\|}$ is responsible for weak tetragonality of the sample.

\begin{figure}[htbp]
  \includegraphics[width=1\columnwidth]{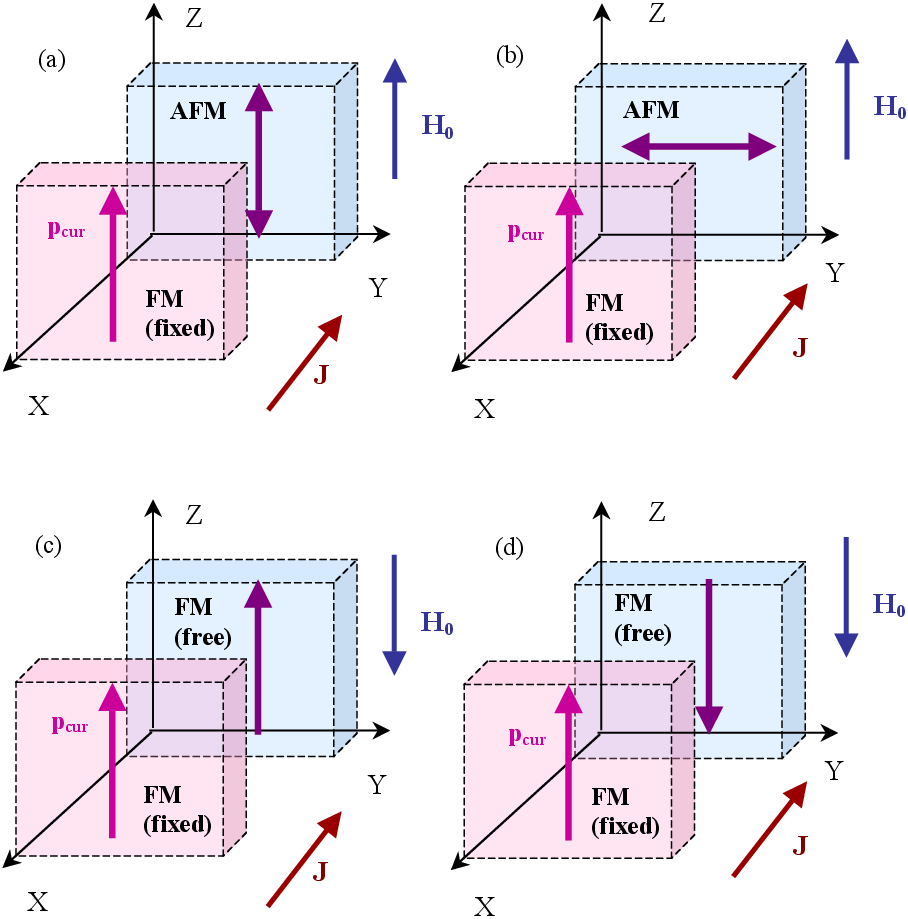}
  \caption{(Color online) \textbf{Stable magnetic configurations with different macroscopic properties.}
In FM/AFM bilayer (a,b) the magnetization ($\|\mathbf{p}_{\rm
cur}$) of FM layer is fixed, AFM vector can be switched from (a)
parallel to (b) perpendicular orientation with respect to
$\mathbf{p}_{\rm cur}$ by application of the external magnetic
field $H_0\ge H_{\rm s-f}$ within an AFM layer. In FM/FM bilayer
(c,d) the magnetization of free layer can be switched from (c)
parallel to (d) antiparallel orientation with respect to
$\mathbf{p}_{\rm cur}$ by the external magnetic field
$\mathbf{H}_0$ applied antiparallel to $\mathbf{p}_{\rm cur}$. In
both cases (c) and (d) the switching can be also induced by
current. \label{fig_geometry_spin_flop}}
\end{figure}

In the absence of field and current a single AFM layer has two
equivalent equilibrium orientations of AFM vector (see
Fig.~\ref{fig_geometry_spin_flop}a, b): $\mathbf{l}\|Z$ and
$\mathbf{l}\|Y$ (as can be easily obtained from minimization of
the magnetic energy (\ref{anisotropy})). Correspondingly, a FM/AFM
bilayer  has two stable configurations\footnote{~In the bilayer
structure the orientation of AFM vector is governed by competition
between the intrinsic magnetic anisotropy (\ref{anisotropy}) and
exchange coupling with the adjacent FM layer. However, in the
typical cases an effective field of exchange coupling is much
smaller than the spin-flop field of AFM and thus can be
neglected.} with $\mathbf{l}\|\mathbf{p}_{\rm cur}$
(Fig.~\ref{fig_geometry_spin_flop}a) and with
$\mathbf{l}\perp\mathbf{p}_{\rm cur}$
(Fig.~\ref{fig_geometry_spin_flop}b). These two configurations
should have different macroscopic properties (e.g., different
magnetoresistance, different exchange bias field, etc.) and in
this sense are analogous to the parallel (P) and antiparallel (AP)
configurations of FM/FM multilayers
(Fig.~\ref{fig_geometry_spin_flop} c,d). In analogy with FM/FM
systems, the reversible switching between the
$\mathbf{l}\|\mathbf{p}_{\rm cur}$ and
$\mathbf{l}\perp\mathbf{p}_{\rm cur}$ states can be achieved by
application of the external magnetic field to the free (in our
case, AFM) layer.

The switching field should be oriented parallel to AFM vector, its
critical value coincides with the spin-flop transition field
$H_{\rm s-f}=2\sqrt{H_{\rm{an}\|}H_E}$ for AFM layer (also
exchange enhanced).

When the current is injected into bilayer, configuration with
$\mathbf{p}_{\rm cur}\|\mathbf{l}$ is still equilibrium, but not
necessarily stable. To find the values of critical current and
field,  we analyze the frequencies of eigen modes of AFM layer
(magnetization of FM layer $\mathbf{p}_{\rm cur}$ and,
correspondingly, the current polarization being fixed).

\subsection{Configuration $\mathbf{H}_0\perp\mathbf{l}$}
 In the crossed initial orientation
$\mathbf{H}_0\perp\mathbf{l}$ (Fig.\ref{fig_geometry_new_2}c) the
linearized equations of motion
 for the generalized coordinates $l_X$, $l_Y$ ($l_Z\approx
2M_0-(l_X^2+l_Y^2)/2$) take the following form:
\begin{eqnarray}\label{motion_long_field}
&&\ddot{l}_X+2\gamma_{\rm
AFM}\dot{l}_X+\left(\omega_X^2+\omega_H^2\right){l}_X+\gamma
H_E\sigma Jl_Y=0,\nonumber\\ &&\ddot{l}_Y+2\gamma_{\rm
AFM}\dot{l}_Y+\omega_Y^2{l}_Y-\gamma H_E\sigma Jl_X=0,
\end{eqnarray}
Here $\gamma_{\rm AFM}\equiv \gamma H_E\alpha_G/2\ll\omega_{X,Y}$
is a damping coefficient that can be estimated from the line-width
of AFM resonance, $\omega_H=\gamma H_0$. The values
$\omega_X=2\gamma\sqrt{(H_{\rm{an}\perp}+H_{\rm{an}\|})H_E}$,
$\omega_Y=2\gamma\sqrt{H_{\rm{an}\|}H_E}$ are the eigen
frequencies of free oscillations in the absence of field and
current that also could be measured in AFM resonance experiments.
It is worth noting that the values of eigen frequencies are
enhanced due to exchange coupling (multiple $H_E$) compared to
analogous values in FM with the same value of anisotropy field.

Equations (\ref{motion_long_field}) describe the case when the
magnetic field is directed along the hard anisotropy axis,
$\mathbf{H}_0\|X$. Configuration with $\mathbf{H}_0\|Y$ (field is
parallel to an easy axis) is treated in an analogous way.

It can be easily seen from (\ref{motion_long_field}) that below
the  critical current
\begin{equation}\label{current_critical_1}
 |J|\le J^{(1)}_{\rm cr}\equiv
\frac{1}{2\gamma H_E\sigma }|\omega^2_X-\omega^2_Y+\omega^2_H|.
\end{equation}
 the eigen modes have linear polarization
and correspond to oscillations of vector $\mathbf{l}$ within $XZ$
or $YZ$ plane (Fig.~\ref{fig_geometry_new_2}c).
  In this case the spin torque
transferred from the current affects the eigen frequencies of spin
excitations,
\begin{eqnarray}\label{frequency_1}
  \Omega^2_{\pm}&=&\frac{1}{2}\left[\omega^2_X+\omega^2_Y+\omega^2_H\right.\nonumber\\
  &  \pm&\left.
\left|\omega^2_X-\omega^2_Y+\omega^2_H\right|
\sqrt{1-\left(\frac{J}{J^{(1)}_{\rm {cr}}}\right)^2} \right],
\end{eqnarray}
 but does not affect the effective damping coefficients (as it
is the case in FM).

It should be stressed that in the absence of external field the
value of critical current depends upon anisotropy
$(\omega^2_X-\omega^2_Y)\propto H_{\rm{an}\perp}$ of the magnetic
interactions within and perpendicular to the film plane. The
magnetic field applied perpendicular to the vector $\mathbf{l}$
enhances (if $\mathbf{H}_0$ is parallel to an easy axis) or
weakens (if $\mathbf{H}_0$ is parallel to a hard magnetic axis of
AFM) the effective anisotropy. So, magnetic field can be used for
control of the critical current. If anisotropy is weak
 ($H_{\rm{an}\perp}\ll H_{\rm{an}\|}$, or, equivalently, $|\omega_X-\omega_Y|\ll \omega_Y$), it can be effectively reduced
with the field whose value is much less than the spin-flop one,
$H_0\ll H_{\rm s-f}$.

Above the critical current, $|J|\ge J^{(1)}_{\rm cr}$,
polarization of free oscillations changes from linear to circular
(elliptic) and STT contributes into the energy dissipation. For
one of two
 modes of free oscillations the current-induced pumping competes with the internal
damping. Starting from the critical value
\begin{equation}\label{current_critical_long_2}
J^{(2)}_{\rm cr}\equiv \sqrt{(J^{(1)}_{\rm
cr})^2+\frac{2\gamma^2_{\rm
AFM}}{\gamma^2H_E^2\sigma^2}\left(\omega^2_X+\omega^2_Y+\omega^2_H\right)},
\end{equation}
the average energy losses per oscillation period are negative
(pumping is greater than damping) and the state with the parallel
alignment of current polarization and AFM vector becomes unstable.

As seen from Eq.(\ref{current_critical_long_2}), the value of
critical current $J^{(2)}_{\rm cr}$ is independent on the
directions of current, field and spin polarization
($\mathbf{p}_{\rm curr}$), in contrast to the threshold current
for FM.

\subsection{Configuration $\mathbf{H}_0\|\mathbf{l}$}
Another type of dynamics is observed in the case when the field is
applied parallel to  $\mathbf{l}$
(Fig.~\ref{fig_geometry_new_2}d). In this case polarization of
eigen modes is circular (or elliptic) even for $J=0$, as follows
from symmetry considerations and from analysis of equations of
motion written in terms of appropriate generalized coordinates
$l_{\pm}=l_X\pm il_Y$:
\begin{widetext}
\begin{equation}\label{motion_short_field}
\ddot{l}_\pm+2\left(\gamma_{\rm AFM}\mp
i\omega_H\right)\dot{l}_\pm
+\left[\frac{1}{2}(\omega_X^2+\omega_Y^2)-\omega^2_H\mp i\gamma
H_E\sigma
J\right]{l}_\pm+\frac{1}{2}(\omega_X^2-\omega_Y^2)l_\mp=0.
\end{equation}
\end{widetext}
 So, oscillations of $\mathbf{l}$  can actively
take up an energy from the current and STT affects the damping
coefficient, not the frequency of oscillations. Instability point
is attained as soon as the spin-polarized current overcomes the
effect of internal friction.

\begin{figure}[htbp]
  \includegraphics[width=1\columnwidth]{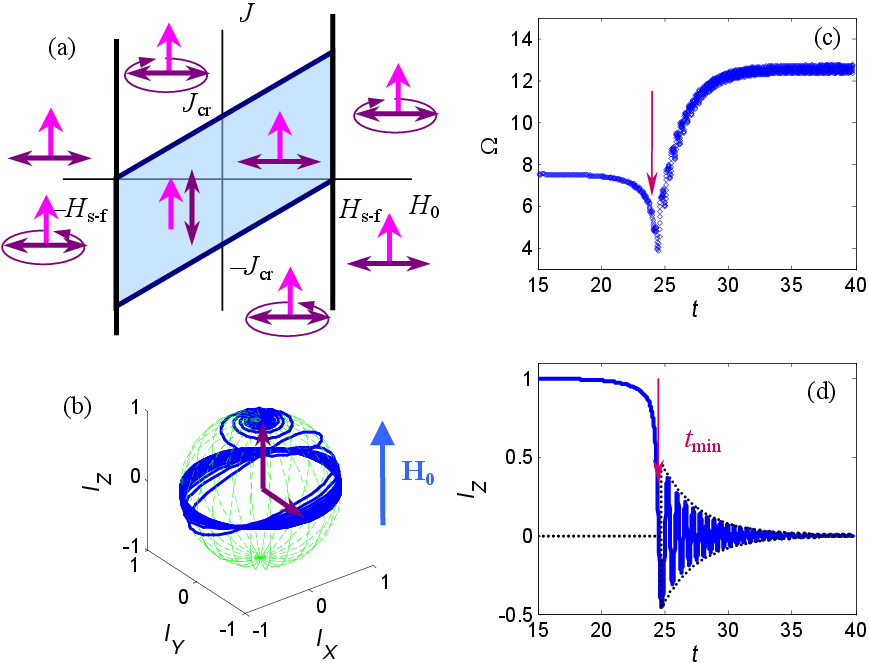}
  \caption{(Color online) \textbf{Dynamics of AFM vector induced by steady
spin-polarized current}. (a) Stability diagram under combined
action of field and current. Static states of the bilayer
(schematically shown with an arrow for $\mathbf{p}_{\rm cur}$ and
double arrow for AFM vector) are stable within the shaded area.
Strong current ($|J|>|J_{\rm cr}|$) induces precession of an AFM
vector around the current polarization $\mathbf{p}_{\rm cur}$,
direction of rotation depends on the sign of $J$. Strong field
($|H_0|>H_{\rm {s-f}}$) induces spin-flop transition combined with
the current-induced precession. (b) 3-D evolution of AFM vector
$\mathbf{l}$ in overcritical regime ($J>J_{\rm cr}$) in the
presence of magnetic field $\mathbf{H}_0$. In the initial state
the AFM vector is slightly misaligned from $Z$ axis. Under the
action of STT the vector $\mathbf{l}$ spirals away from $Z$ axis
with a steadily-increasing precession angle and the angular
frequency $\Omega$. Final state corresponds to precession with the
stable frequency within $XY$ plane ($l_Z=0$). (c), (d) Time
dependence of the angular frequency $\Omega$ and $l_Z$ projection.
Arrows indicate the moment, $t_{\rm min}$, at which monotonic
decrease of $l_Z$ switches to the decaying oscillations. Envelope
(dash line) corresponds to relaxation ($\propto \exp(-\gamma_{\rm
AFM}t)$) caused by internal damping. \label{freq_steady_green}}
\end{figure}

Fig.~\ref{freq_steady_green}a) shows the field--current stability
diagram for the case of isotropic AFM ($H_{\rm{an}\perp}=0$ and,
correspondingly, $\omega_X=\omega_Y$). Within the shaded area
\begin{equation}\label{stability_inequalities}
  \left|J-\frac{2\gamma_{\rm AFM}H_0}{\sigma H_E}\right|\le \frac{2\gamma_{\rm AFM}H_{\rm
s-f}}{\sigma H_E},\quad |H_0|\le H_{\rm s-f},
\end{equation}
 the static state with $\mathbf{l}\|\mathbf{p}_{\rm cur}\|Z$ is stable. Above the
critical value, $|J|\ge |J_{\rm cr}|$, where
\begin{equation}\label{critical_current_2}
  J_{\rm cr}\equiv \frac{2\gamma_{\rm AFM}}{\sigma H_E}\left(H_0+H_{\rm s-f}{\rm{sign}}J
\right),
\end{equation}
 the current may keep up a stable rotation of AFM vector around
$\mathbf{p}_{\rm cur}$ (Fig.~\ref{freq_steady_green}b). Sign
reversal of STT (resulted from the reversal of either direction of
current, $J\rightarrow -J$, or direction of polarization,
$\mathbf{p}_{\rm cur}\rightarrow-\mathbf{p}_{\rm cur}$) gives rise
to rotation in opposite direction. If the field value is greater
than spin-flop field, $|H_0|\ge H_{\rm s-f}$, the state with
$\mathbf{l}\|\mathbf{H}_0$ is unstable even in the absence of
current and in the final state the AFM vector is perpendicular to
$\mathbf{H}_0$ and $\mathbf{p}_{\rm cur}$. These results also keep
true for small but nonzero anisotropy $H_{\rm{an}\perp}$.

The current-induced precession of $\mathbf{l}$ vector is also
stable in the ``high-field'' region, $|H_0|\ge H_{\rm s-f}$.
However, detailed analysis of the dynamical phases and transition
lines in this region is out of scope of this paper.

Some features of the current-induced instability in configuration
with $\mathbf{l}\|\mathbf{H}_0$ are similar to those observed in
FM/FM bilayers. First, in both cases the stability region is
defined by an internal friction which stands up against the
current-induced rotations \cite{Kiselev:2003}. Second, the value
of critical current linearly depends on the field
\cite{Grollier:2003,Gulyaev:2005}. Thus, application of the
magnetic field results in variation of the critical current and
opens a possibility to reduce $J_{\rm cr}$, as seen from
Fig.~\ref{freq_steady_green}a.

\begin{figure}[htbp]
  \includegraphics[width=1\columnwidth]{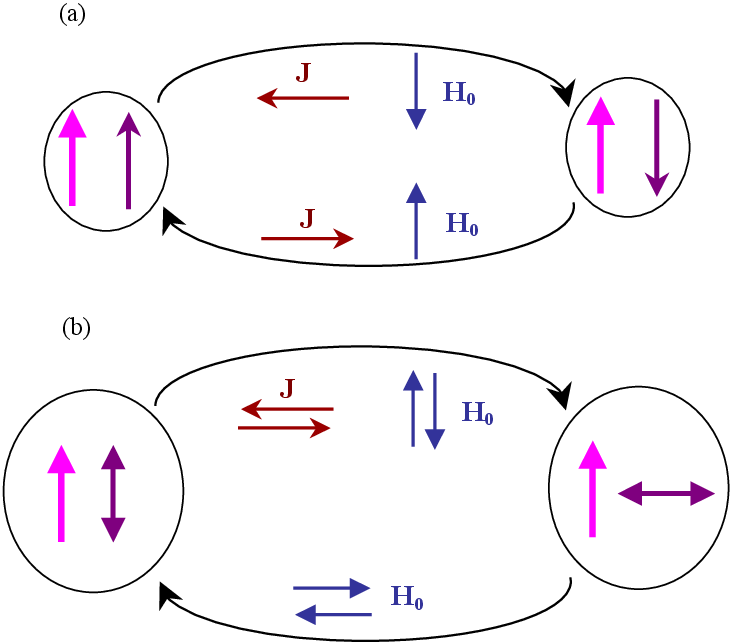}
  \caption{(Color online) \textbf{Switching between the different configurations of FM/FM (a) and FM/AFM (b)
  bilayers.} Magnetization of the fixed layer is shown with magenta
  (thick) arrow, that of the free layer with violet (thin arrow), double arrow shows orientation of AFM vector. (a) Switching between P and AP states can be
  achieved by the field or current applied in two opposite
  directions. (b) Transition from parallel to perpendicular
  configuration can be induced by current (arbitrary direction)
  and field applied along initial orientation of AFM vector.
  Transition from perpendicular to parallel configuration can be
  induced by the field only.
 \label{fig_switching}}
\end{figure}

On the other hand, there are still few principal differences
between FM/FM and FM/AFM bilayers listed below.
\begin{itemize}
  \item In the FM/FM bilayer (one FM layer being fixed) switching
  between P and AP states can be achieved by application of \textit{either} field or
  current (Fig.~\ref{fig_switching}a). In contrast, in the FM/AFM bilayer switching between $\mathbf{l}\|\mathbf{p}_{\rm cur}$ and $\mathbf{l}\perp\mathbf{p}_{\rm
  cur}$ states can be achieved by a \textit{combined} application of field and current (Fig.~\ref{fig_switching}b). Namely, the current
  induces
  transition only from $\mathbf{l}\|\mathbf{p}_{\rm cur}$ to $\mathbf{l}\perp\mathbf{p}_{\rm
  cur}$ because the last state is stable in the presence of current. To reverse $\mathbf{l}$ vector back to $\mathbf{l}\|\mathbf{p}_{\rm
  cur}$ configuration, one needs to apply an external field $H_0\ge
  H_{\rm s-f}$ parallel to $\mathbf{l}$ (spin-flop transition).
  \item In the FM/FM bilayer the direction of current (from fixed
  to free layer or opposite) is important,
  P $\rightarrow$ AP and AP $\rightarrow$ P transitions take place
  at opposite directions of current (Fig,~\ref{fig_switching}a). In contrast, in the FM/AFM
  bilayer destabilization of $\mathbf{l}\|\mathbf{p}_{\rm cur}$
  state takes place irrespective of the current direction (Fig.~\ref{fig_switching}b).
  However, an external magnetic field removes such a degeneracy.
  \item The bilayers with AFM should show exchange reduction of
  the critical current compared to FM/FM bilayers providing that
  the free FM and AFM layers have the same magnetic resonance frequencies (or anisotropy field of FM is close to spin-flop field of
  AFM) and the same quality factor ($=\omega/\gamma_{\rm AFM}$), as can be seen from Eq.~(\ref{critical_current_2}).
\end{itemize}

\section{Dynamics in overcritical regime}
A FM layer subjected to the direct spin-polarized current shows
one interesting effect -- stable precession of magnetization with
the angular frequency close to the frequency of spin-wave mode
\cite{Kaka:2005,Slavin:2009}. To find out whether such an effect
could be observed in AFM, we consider in details the dynamics of
AFM vector in overcritical regime ($|J|>|J_{\rm cr}|$) assuming
that $\mathbf{p}_{\rm cur}\|\mathbf{H}_0\|Z$.

We use the standard parametrization of AFM vector with the
spherical angles $\theta$ and $\varphi$,
$l_X=2M_0\sin\theta\cos\varphi$, $l_Y=2M_0\sin\theta\sin\varphi$,
$l_Z=2M_0\cos\theta$, to deduce the following dynamic equations:
\begin{widetext}
\begin{eqnarray}\label{dynamics_angles_2}
&&\ddot {\theta}+2\gamma_{\rm AFM}
\dot{\theta}+\sin\theta\cos\theta\left[\omega^2_Y-
\left(\dot{\varphi}-\omega_H\right)^2
+\frac{\omega^2_X-\omega^2_Y}{2}(1+\cos2\varphi)
-\frac{\omega^2_Y-\omega_H^2}{4}\sin^2\theta\left(7+\cos4\varphi\right)\right]=0,\nonumber\\
&&\frac{d}{dt}\left[\left(\dot{\varphi}-\omega_H\right)\sin^2\theta\right]+
\sin^2\theta\left[2\gamma_{\rm AFM}\dot{\varphi}-\gamma
H_EJ\sigma+{\frac{\omega^2_Y-\omega^2_X}{2}\sin2\varphi+\frac{\omega^2_Y-\omega_H^2}{4}\sin^2\theta\sin4\varphi}
 \right]=0.
\end{eqnarray}
\end{widetext}
As it was already mentioned, an AFM under consideration is an
oscillator with the high quality factor ($\omega_{X,Y}\gg
\gamma_{\rm AFM}$). In other words, energy dissipation takes place
on the time scale much greater than the characteristic period of
free oscillations. In this case for analytical treatment of
Eqs.~(\ref{dynamics_angles_2}) one can apply the asymptotic method
of rapidly rotating phase originated by Bogolyubov and
Mitropolskii \cite{Bogoluybov:1998}.

\begin{figure}
  \includegraphics[height=.3\textheight]{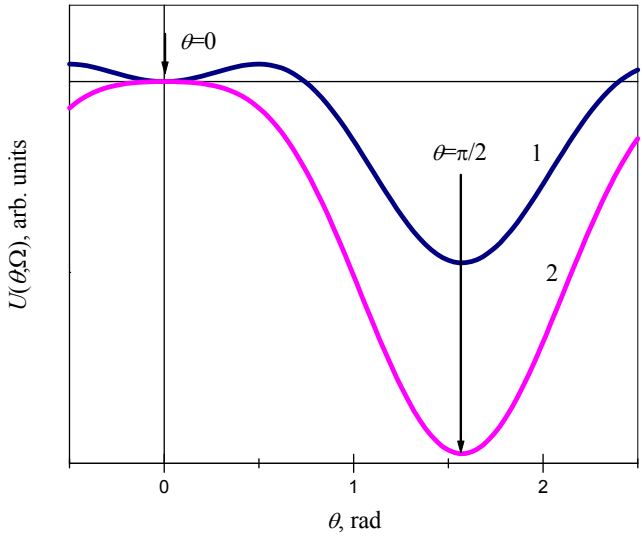}
  \caption{Profile of the effective potential well (\ref{potential_well}), arbitrary units. 1 -- $\Omega\le \Omega_0$, 2 -- $\Omega > \Omega_0$. Arrows indicate static solution ($\theta=0$) unstable in overcritical regime and stable stationary solution ($\theta=\pi/2$). \label{fig_potential_well}}
\end{figure}

According to this method, the motion of AFM vector is decomposed
into rapid rotation with the frequency $\Omega\propto\omega_{X,Y}$
and slow variation of amplitude and frequency with the
characteristic time scale $\propto1/\gamma_{\rm AFM}$. In the
simplest case of isotropic AFM ($H_{\rm{an}\perp}=0$, or
$\omega_X=\omega_Y$) the only rapid variable is
$\varphi=\Omega(t)t$. Equations for slow variables $\Omega(t)$ and
$\theta(t)$ ($\dot{\Omega}, \dot{\theta}\ll \Omega$) are obtained
from (\ref{dynamics_angles_2}) by averaging over the period of
rotation:
\begin{eqnarray}\label{dynamics_angles_3}
&&\ddot {\theta}+2\gamma_{\rm AFM} \dot{\theta}\nonumber\\
&&+\sin\theta\cos\theta\left[\omega^2_0-
\left(\Omega-\omega_H\right)^2 -\frac{7}{4}
\omega^2_0\sin^2\theta\right]=0,\\
&&\frac{d}{dt}\left[\left(\Omega-\omega_H\right)\sin^2\theta\right]+
\sin^2\theta\left(2\gamma_{\rm AFM}\Omega-\gamma H_EJ\sigma
\right)=0.\nonumber
\end{eqnarray}
If, in addition, $\dot{\Omega}\ll \dot{\theta}$, the first of
equations (\ref{dynamics_angles_3}) describes 1D motion (dynamic
variable $\theta$) in a potential well (see
Fig.~\ref{fig_potential_well})
\begin{equation}\label{potential_well}
U(\theta;
\Omega)=\frac{1}{2}\sin^2\theta\left[\omega^2_0-\left(\Omega-\omega_H\right)^2-\frac{7}{4}
\omega^2_0\sin^2\theta\right],
\end{equation}
with the friction defined by coefficient $\gamma_{\rm AFM}$. The
second of Eqs.~(\ref{dynamics_angles_3}) describes the
current-induced variation of both variables $\theta$ and $\Omega$.

Equations (\ref{dynamics_angles_3}) have two interesting
solutions. The first one, corresponds to the circular polarized
free oscillations of AFM vector with an amplitude
$\theta=\theta_0\ll 1$ and eigen frequency $\Omega_0\equiv
\omega_X+\omega_H$. However, in overcritical regime an amplitude
$\theta_0$ growth with an increment proportional to the offset
from the critical current value:
\begin{equation}\label{time_dependence_theta_1}
\frac{1}{\tau}\equiv\gamma_{\rm AFM}\left|\frac{J-J_{\rm
cr}}{J_{\rm cr}}\right|\left(1+\frac{H_0}{H_{\rm s-f}}\right).
\end{equation}

The second solution with $\theta=\pi/2$ corresponds to steady
rotation of AFM vector in $XY$ plane ($l_Z=0$) with the angular
frequency $\Omega_\infty=(J/J_{\rm cr})\Omega_0$. Energy
dissipation per period of rotation is zero, due to the pretty
balance between the magnetic damping and current-induced pumping.
This solution is stable when $|J|>|J_{\rm cr}|$, as can be seen
from analysis of the potential $U(\theta; \Omega)$. Small
deviations of AFM vectors from $XY$ plane ($|\theta-\pi/2|\ll 1$)
relax due to internal friction as
\begin{equation}\label{final_stage}
  \cos\theta\propto e^{-\gamma_{\rm AFM}t}\cos(\Omega_\theta
  t+\psi),
\end{equation}
where $\psi$ is a phase that depends upon initial conditions and
\begin{equation}\label{frequency_theta}
  \Omega_\theta=\sqrt{\Omega_\infty^2-2\Omega_\infty\omega_H+\frac{3}{4}\left(\omega_0^2-\omega_H^2\right)}.
\end{equation}
So, the state of steady precession is approached during the time
$1/\gamma_{\rm AFM}$ that depends upon the internal magnetic
damping.

To illustrate all the described peculiarities of AFM dynamics in
the presence of spin-polarized current we solve the original
Eqs.~(\ref{dynamics_angles_2})
 numerically with the initial conditions $\theta=\theta_0=0.001$, $\varphi=\pi/2$,
$\dot{\theta}=0$ and $\dot{\varphi}=\Omega_0$. In other words, at
$t=0$ an AFM vector deflects from equilibrium orientation
$\mathbf{l}\|Z$ through the small angle $\theta_0\ll 1$ within the
$ZY$ plane. Initial velocity corresponds to that mode of free
oscillations which is unstable for the chosen current direction.
For calculations we used the following dimensionless values:
$\omega_X=\omega_Y=6.28$ and $\gamma_{\rm AFM}=0.314$ (that
corresponds to the quality factor 20). Time unit equals to the
period of free oscillations in the absence of field and current.

Fig.~\ref{freq_steady_green}b shows a typical trajectory of AFM
vector (normalized to a unit length) in the presence of steady
current $J=2.5 J_{\rm cr}<0$ and field $H=0.2H_{\rm s-f}$. With
the described initial conditions, the motion of $\mathbf{l}$
vector starts as a rotation around $Z$ axis with the eigen
frequency $\Omega_0$ of free oscillations. Due to the energy
pumping from STT, an amplitude of oscillations ($\mathbf{l}$
projection on $XY$ plane) slowly increases with an increment
$\tau$ (see Eq.~(\ref{time_dependence_theta_1})).

The final state ($t\rightarrow\infty$) corresponds to the above
described steady rotation of AFM vector in $XY$ plane ($l_Z=0$)
with the angular frequency $\Omega_\infty$. In analogy with FM,
such a precessional state of an AFM layer can be a source of spin
waves. In contrast to FM, the angular frequency $\Omega_\infty$ is
proportional to the current value. The absolute value of
$\Omega_\infty$ is greater than the characteristic spin-wave
frequency ($\propto\Omega_0$) which in AFM can range THz values
(e.g., for bulk FeMn the energy gap is 7 meV \cite{Endoh:1973}
that corresponds to linear frequency $\propto$2~THz). So, FM/AFM
bilayer can be considered as a potential emitter of high frequency
spin waves.

\begin{figure}[htbp]
\includegraphics[width=1\columnwidth]{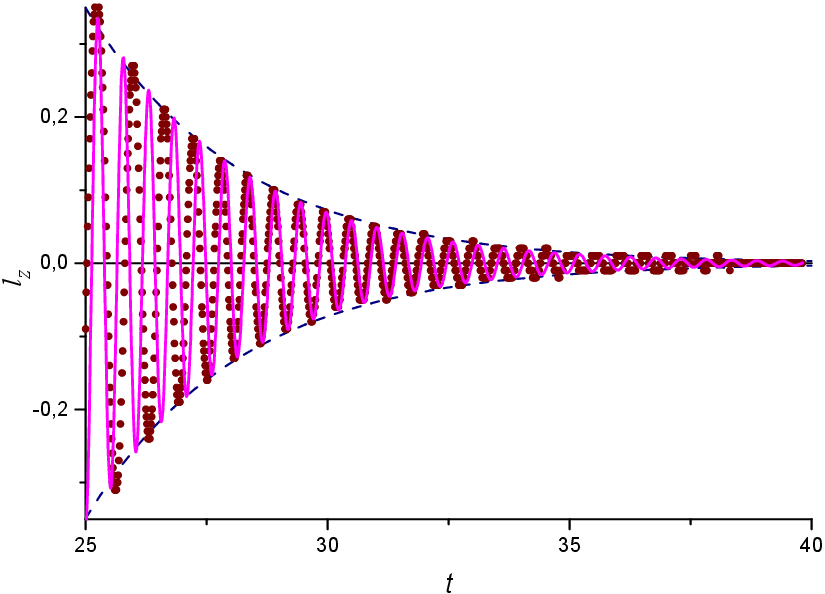}
  \caption{(Color online) Relaxation dynamics of slow amplitude $l_Z$ at $t\ge t_{\rm min}$. Points -- numerical simulation, solid line -- approximation according to Eq.~(\ref{final_stage}), dashed line -- envelope $\propto \exp(-\gamma_{\rm AFM}t)$.\label{oscillations_end}}
\end{figure}

Figs.~\ref{freq_steady_green}c and \ref{freq_steady_green}d
illustrate the time evolution of the rotation frequency $\Omega$
and the component $l_Z$ between the initial and final states.
 Due to nonlinear effects, deflection of
$\mathbf{l}$ from the initial direction is accompanied by decrease
of $\Omega$. At a certain moment $t=t_{\rm min}$ (shown by arrow
in Figs.~\ref{freq_steady_green} c,d) a monotonic decrease of
$l_Z$ changes into decaying oscillations around the average value
$l_Z=0$. Relaxation of $\mathbf{l}$ to $XY$ plane is due to
internal damping and follows the law $\propto \exp(-\gamma_{\rm
AFM}t)$ (see Eq.~(\ref{final_stage}) and envelope in
Fig.~\ref{freq_steady_green}d). The details of relaxation to the
precessional state are shown in  Fig.~\ref{oscillations_end} where
we compare an exact solution (points) of (\ref{dynamics_angles_2})
with the asymptotic form (solid line) calculated from the
expressions (\ref{time_dependence_theta_1}), (\ref{final_stage}).

In our simulations of AFM dynamics we found that not only $l_Z$,
but the rotation frequency $\Omega$, energy dissipation rate and
the effective potential energy averaged over a period of rotation
($=2\pi/\Omega$) has an extremum at the moment $t=t_{\rm min}$.
This means that the system passes through the crossing between two
attraction points in the phase space. So, we interpret the time
interval $t_{\rm min}$ as a switching time between two stable
states of the FM/AFM bilayer (see
Figs.~\ref{fig_geometry_spin_flop} a,b, and \ref{fig_switching}
b).
\begin{figure}[htbp]
\includegraphics[width=1\columnwidth]{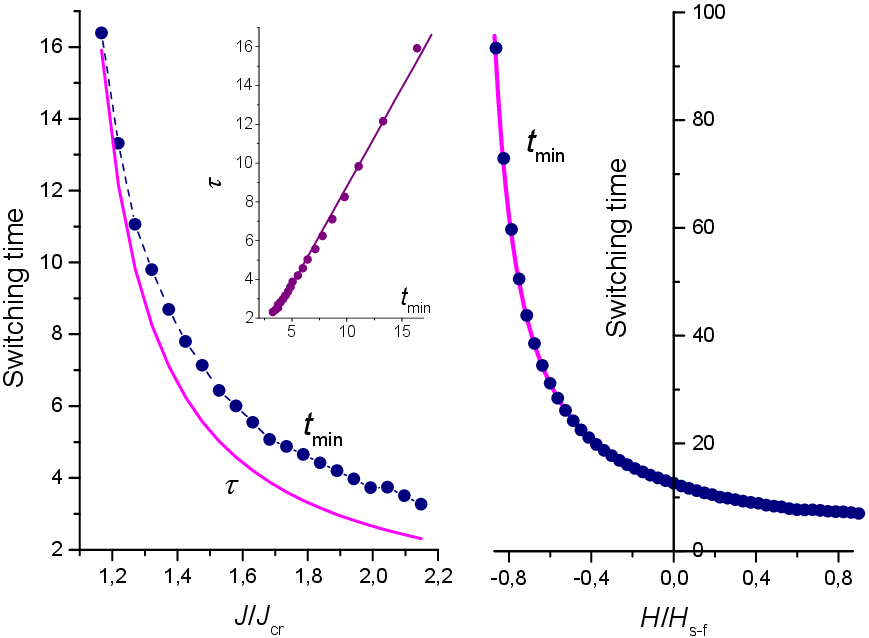}
  \caption{(Color online) \textbf{Switching time from parallel to perpendicular configurations shown in Fig.~\ref{fig_geometry_spin_flop} a,b.} (a) Current dependence of the calculated switching  time, $t_{\rm min}$, and the effective pumping coefficient $\tau\propto|J-J_{\rm cr}|^{-1}$, external magnetic field $H_0=0.2H_{\rm sf}$ is parallel to $Z$.
 Inset shows correlation between $\tau$ and $t_{\rm min}$. (b) Field dependence of switching time $t_{\rm min}$ at fixed bias current (points). Solid line shows approximation $\propto
 |H_{\rm sf}+H_0|^{-1}$.\label{time_vs_current_field}}
\end{figure}
The exact value of $t_{\rm min}$ depends upon the initial
deflection $\theta_0$ of AFM vector from $Z$ axis (or, in other
words, from the amplitude of spontaneous fluctuations and, hence,
from temperature). However, the current and field behaviour of
$t_{\rm min}$ is the same for different initial conditions and
correlates with the current (Fig.~\ref{time_vs_current_field}a)
and field (Fig.~\ref{time_vs_current_field}b) behavior of the
characteristic time $\tau$ of destabilization
(\ref{time_dependence_theta_1}). Though the absolute values of
$t_{\rm min}$ and $\tau$ are different, they show good correlation
(see inset in Fig.~\ref{time_vs_current_field}a) in rather wide
range of current values. Both values decrease moving far from the
stability point $J_{\rm cr}$. This opens a way to diminish the
 switching time by increasing the current value or by decreasing the
critical current with the magnetic field. 

\section{Controlable switching of AFM state}
\begin{figure}[htbp]\centering
   \mbox{\epsfig{file=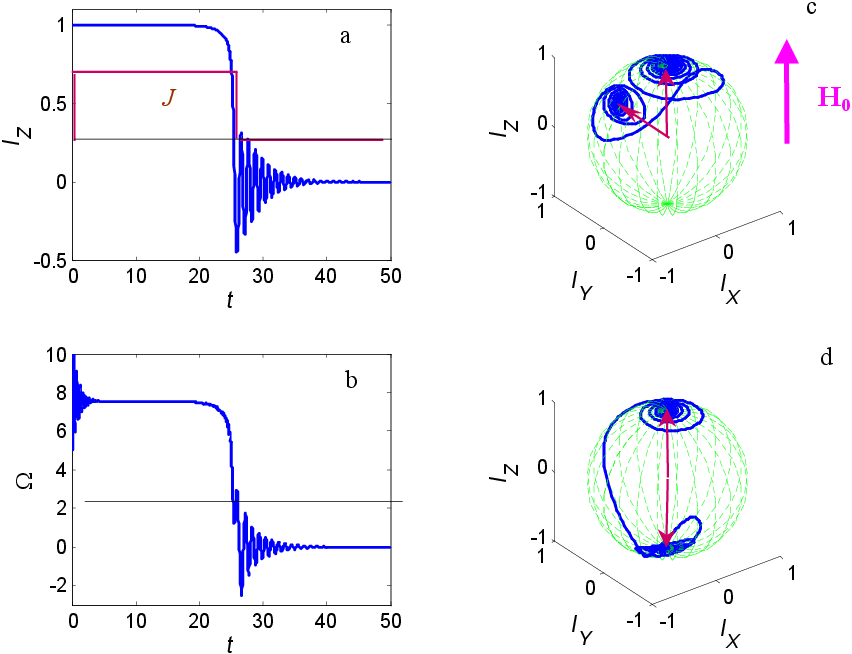,width=0.8\columnwidth}}
    \caption{(Color online) \textbf{Dynamics of AFM vector induced by a pulse of spin-polarized
current.} (a) Time dependence of $l_Z$ (thick or blue line) under the
rectangular current pulse (thin or red line). Pulse duration close to
$t_{\rm min}$ is enough to ensure maximal deflection of $l_Z$ from
the initial value. (b) High frequency rotation of AFM vector
persists as long as current induces STT, as seen from time
dependence of $\Omega$. Just after the current is switched off
the frequency of rotation goes down to zero and AFM vector lies down
to $XY$ plane. Relaxation time ($=1/\gamma_{\rm AFM}$) of both
$\Omega$ and $l_Z$ is due to internal damping. (c) Magnetic field
applied parallel to the initial orientation of $\mathbf{l}$ induces
90$^\circ$ switching, the final state of AFM vector is parallel to $XY$ plane. (d) In
isotropic AFM in the absence of field a spin-polarized current may
induce 180$^\circ$
switching.\label{current_pulse_green}}\end{figure}

From practical point of view it is important to achieve a
controllable switching between the different equilibrium states of
the FM/AFM bilayer (say, $\mathbf{l}\|\mathbf{p}_{\rm cur}$
$\rightarrow$ $\mathbf{l}\perp \mathbf{p}_{\rm cur}$ or
$\mathbf{l}\|-\mathbf{p}_{\rm cur}$) using the current pulses of
minimal duration and amplitude. We investigate dynamics of AFM
vector under the rectangular current pulses (schematically shown
by red line in Fig.\ref{current_pulse_green}) of different
duration and amplitude in overcritical regime ($|J|>|J_{\rm
cr}|$). If pulse duration is below $t_{\rm min}$, AFM vector
returns back to its initial state after the current is switched
off.

Fig.~\ref{current_pulse_green} demonstrates the switching processes
initiated by the current pulse $J=2.5J_{\rm cr}$ with the duration
slightly greater than $t_{\rm min}$. The chosen pulse duration
ensures maximum deflection of AFM vector from the initial
direction as seen from time dependence of $l_Z$
(Fig.~\ref{current_pulse_green} a). After the current is switched
off, the AFM vector relaxes to the final static state within
$XY$ plane through the damping oscillations during the time
$1/\gamma_{\rm AFM}$. Regular rotation around $Z$ axis supported
by current also vanishes with the end of current pulse, as seen
from $\Omega$ behaviour (Fig.~\ref{current_pulse_green}b). The
final orientation of AFM vector is parallel to one of the easy
axes within $XY$ plane (in the presence of field,
Fig.~\ref{current_pulse_green}c) or can be also anitiparallel to
the initial $\mathbf{l}$ direction (180$^\circ$ switching to $Z$ easy axis) in the
absence of external field (Fig.~\ref{current_pulse_green}d). Due
to degeneracy, the final state is very sensitive to the
initial conditions and pulse duration and can be predicted only
statistically.

\section{Conclusions}

In summary, we studied the dynamics of AFM layer in the presence
of spin-polarized current and external magnetic field.

On the basis of a simple model of slightly tetragonal AFM with two magnetic sublattices we demonstrated the following features of current-induced behavior in the FM/AFM bilayer.
\begin{enumerate}
\item Spin transfer torque induces the loss of stability in the  FM/AFM system if  AFM vector is parallel to FM magnetization.  Configuration with AFM vector perpendicular to FM magnetization is stable in the presence of current and field (in accordance with the prediction made in Ref.\onlinecite{Haney:2007(2)}). This means that the high density current can induce reorientation of AFM vector if an angle between $\mathbf{l}$ and $\mathbf{p}_{\rm cur}$  differs from $\pi/2$.
\item Such a current can also induce a stable precession of AFM vector in the plane perpendicular to FM magnetization. The frequency of precession is of the order of frequency of free oscillations and linearly depends on current.
\item The value of critical current can be tuned by application of the external magnetic field. The value of switching time can be tuned by both field and current.
\end{enumerate}
We anticipate the same features in noncollinear AFM  with three (like IrMn$_3$ and Mn$_3$NiN) and more (like FeMn) magnetic sublattices.

On the other hand, our model predicts irreversible current-induced switching between the parallel and perpendicular orientations of  $\mathbf{l}$ and  $\mathbf{p}_{\rm cur}$. This result is the consequence of the chosen tetragonal magnetic anisotropy. In the case when an angle between easy axes of AFM differs from $\pi/2$ (like in NiO or FeMn), the current-induced switching seems to be possible between all the configurations of $\mathbf{l}$ and  $\mathbf{p}_{\rm cur}$ as long as $\mathbf{l}$ has a nonzero projection on $\mathbf{p}_{\rm cur}$.

\begin{figure}[htbp]\centering
   \mbox{\epsfig{file=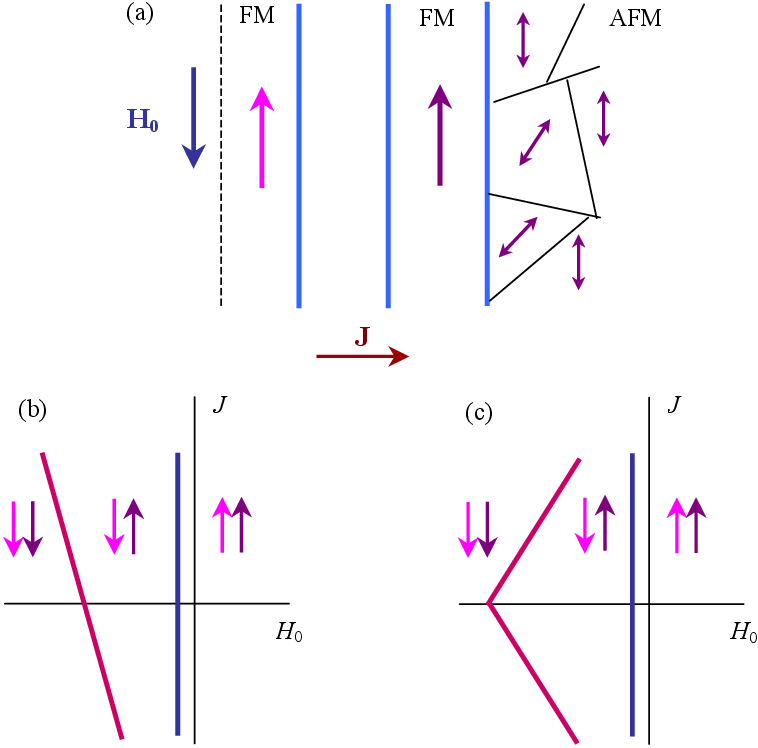,width=0.8\columnwidth}}
    \caption{(Color online) \textbf{Current dependence of the exchange bias field.} (a) Exchange bias
     spin-valve structure with multidomain AFM layer. AFM vector in some of domains deflects from the initial orientation due to STT. (b) The shift of bias for typical spin transfer (AFM layer is not affected by current). (c) The shift of bias in the case when spin torque is transferred to AFM layer.  \label{fig_bias}}\end{figure}

The described response of AFM vector to electron current may
change the properties of the pinned layer. Due to the weak but
nonzero exchange coupling between AFM and FM layers, reorientation
or precession of AFM vector results in variation of the exchange
bias field and, consequently, gives rise to the shift of switching
fields of spin-valve. We illustrate this effect qualitatively in
Fig.~\ref{fig_bias}. Consider a typical spin-valve with the pinned
FM layer. Suppose, an AFM layer is inhomogeneous (multidomain) and
high density current gives rise to reorientation of AFM vector in
some of domains (Fig.~\ref{fig_bias}a). The ratio of the rotated
domains is proportional to the integral current. On the other
hand, reorientation of some AFM domains results in diminishing of
the exchange bias field that keeps FM magnetization of the pinned
layer. Variation of the bias field is also proportional to the
ratio of the rotated AFM domains. So, in the presence of current
the critical field at which magnetization of the pinned layer is
reversed decreases linearly. However, linear shift of the bias
field can be also induced by STT between FM layers.  If AFM layer
is not affected by spin transfer torque, the stability region of
AP configuration increases for one current direction and
diminishes for an opposite, as shown in Fig.~\ref{fig_bias}b. On
the contrary, if spin torque is transferred to an AFM layer and is
not transferred between two FM layers, the stability region of
antiparallel configuration of FM layers diminishes for any current
direction (Fig.~\ref{fig_bias}c). The last type of current
dependence (among others) was observed in the experiments
Ref.\onlinecite{Basset08}. Decrease of the exchange bias field
irrespective of current direction was also observed in
Ref.\onlinecite{Dai08}.

Linear shift of the bias field induced by the current
 was observed  in nanopillars\cite{Urazhdin:2007} that included coupled
permalloy (FM) and FeMn (AFM) layers. In these experiments a
combined application of the magnetic field and high-density
current resulted in an increase of the exchange bias field  from
-100 to 100~Oe.

Another evidence of STT effects in AFM could be found from the
detailed analysis of the field/current dependence of
magnetoresistance, as it was done in Refs.\onlinecite{Tsoi:2007,
tang:122504, Dai08, Tsoi-2008}. Magnetoresistance of spin valve
should depend on the angle between FM and AFM vectors (in addition
to the dependence from mutual orientation of FM vectors in
``free'' and ``pinned'' layers) and can change due to the
current-induced switching  of AFM vector.


\appendix
\section{Relation between magnetization and AFM vector}\label{A}
\renewcommand{\theequation}{A. \arabic{equation}}
\setcounter{equation}{0} Small macroscopic magnetization
$|\mathbf{m}|\ll |\mathbf{l}|$ of AFM layer can be excluded from
Eq.(\ref{Landau-Lifshitz_2a}) in the following way
\cite{Bar-june:1979E,Ivanov:2009}. Free energy of AFM layer is
modeled as
\begin{equation}\label{energy_total}
  w=\frac{H_E}{4M_0}\mathbf{m}^2+w_{\rm an}-\mathbf{H}_0\mathbf{m},
\end{equation}
where $H_E$ is the spin-flip field of the exchange nature, $w_{\rm
an}$ is anisotropy energy, $\mathbf{H}_0$ is external magnetic
field. In assumption that $H_EM_0\gg w_{\rm an}$ (or,
equivalently, $H_E\gg H_L$) and $H_E\gg H_0$ (strong exchange
coupling), the effective magnetic field is expressed as follows:
\begin{equation}\label{effective field}
  \mathbf{H}_M\equiv-\frac{\partial w}{\partial \mathbf{m}}=\mathbf{H}_0-\frac{H_E}{2M_0}\mathbf{m}.
\end{equation}
We substitute (\ref{effective field}) into
(\ref{Landau-Lifshitz_2a}), neglect dissipative terms (with
$\alpha_G$ and $J$) and get
\begin{equation}\label{inter_mangetization}
  \dot{\mathbf{l}}=\gamma\left([\mathbf{H}_0\times
\mathbf{l}]-\frac{H_E}{2M_0}[\mathbf{m}\times\mathbf{l}]\right).
\end{equation}
To obtain an explicit expression (\ref{magnetization}) for
$\mathbf{m}$ we multiply both sides of
Eq.(\ref{inter_mangetization}) by the vector $\mathbf{l}$ and take
into account that
$[\mathbf{l}\times[\mathbf{m}\times\mathbf{l}]=\mathbf{m}\mathbf{l}^2\approx4M_0^2\mathbf{m}$.
The last relation follows from the fact that below the N\`{e}el
temperature both vectors $\mathbf{l}$ and $\mathbf{m}$ are bound
by the constraint $\mathbf{l}^2+\mathbf{m}^2=4M_0^2$,
$(\mathbf{l},\mathbf{m})=0$ (that is equivalent to the requirement
$\vert\mathbf{M}_1\vert=\vert\mathbf{M}_2\vert=M_0$).
\section{Lagrange and Rayleigh functions for
antiferromagnet}\label{B}
\renewcommand{\theequation}{A. \arabic{equation}}
The Lagrange function (\ref{Lagrangian}) and the Rayleigh function
(\ref{Relay}) are selected in such a way that to fulfill the
following requirements:
\renewcommand{\theenumi}{\roman{enumi}}
\renewcommand{\labelenumi}{{\it{\theenumi}})}
\begin{enumerate}
  \item The dynamic Eqs.(\ref{motion_equation}) are the Euler-Lagrange
  equations of the second kind with dissipative forces:
  \begin{eqnarray}\label{euler_equation}
&&\left[\frac{d}{dt}\frac{\partial \mathcal{L}_{\rm AFM}}{\partial
\dot{\mathbf{l}}}\times\mathbf{l}\right]-\left[\frac{\partial
\mathcal{L}_{\rm AFM}}{\partial
\mathbf{l}}\times\mathbf{l}\right]\nonumber\\
&&=-\left[\frac{\partial\mathcal{R}_{\rm AFM}}{\partial
\dot{\mathbf{l}}}\times\mathbf{l}\right].\end{eqnarray}
  \item The effective potential energy in Lagrange function
  coincides with the magnetic anisotropy energy
  (\ref{anisotropy}).
  \item Rayleigh function is related with the rate of energy
  losses (\ref{energy_losses_2}) according to Eq.(\ref{energy_losses}).
\end{enumerate}
Expression for energy losses is obtained from
(\ref{energy_losses_2}) by substitution
$\mathbf{M}_1=-\mathbf{M}_2=\mathbf{l}/2$. Contribution from
$\mathbf{m}$-depending terms into Rayleigh function is small and
so, is neglected.

Lagrange approach makes it possible to account for the constraint
$|\mathbf{l}|=2M_0$ (valid far below the N\`{e}el point) by
appropriate choice of two generalized coordinates $q_k$ ($k=1,2$)
instead of three components of vector $\mathbf{l}$, as described
in the paper.


\begin{thebibliography}{10}%
\makeatletter
\providecommand \@ifxundefined [1]{%
 \ifx #1\undefined \expandafter \@firstoftwo
 \else \expandafter \@secondoftwo
\fi
}%
\providecommand \@ifnum [1]{%
 \ifnum #1\expandafter \@firstoftwo
 \else \expandafter \@secondoftwo
\fi
}%
\providecommand \enquote [1]{``#1''}%
\providecommand \bibnamefont  [1]{#1}%
\providecommand \bibfnamefont [1]{#1}%
\providecommand \citenamefont [1]{#1}%
\providecommand\href[0]{\@sanitize\@href}%
\providecommand\@href[1]{\endgroup\@@startlink{#1}\endgroup\@@href}%
\providecommand\@@href[1]{#1\@@endlink}%
\providecommand \@sanitize [0]{\begingroup\catcode`\&12\catcode`\#12\relax}%
\@ifxundefined \pdfoutput {\@firstoftwo}{%
 \@ifnum{\z@=\pdfoutput}{\@firstoftwo}{\@secondoftwo}%
}{%
 \providecommand\@@startlink[1]{\leavevmode}%
 \providecommand\@@endlink[0]{}%
}{%
 \providecommand\@@startlink[1]{%
  \leavevmode
  \pdfstartlink
   attr{/Border[0 0 1 ]/H/I/C[0 1 1]}%
   user{/Subtype/Link/A<</Type/Action/S/URI/URI(#1)>>}%
  \relax
 }%
 \providecommand\@@endlink[0]{\pdfendlink}%
}%
\providecommand \url  [0]{\begingroup\@sanitize \@url }%
\providecommand \@url [1]{\endgroup\@href {#1}{\urlprefix}}%
\providecommand \urlprefix [0]{URL }%
\providecommand \Eprint[0]{\href }%
\@ifxundefined \urlstyle {%
  \providecommand \doi [1]{doi:\discretionary{}{}{}#1}%
}{%
  \providecommand \doi [0]{doi:\discretionary{}{}{}\begingroup
  \urlstyle{rm}\Url }%
}%
\providecommand \doibase [0]{http://dx.doi.org/}%
\providecommand \Doi[1]{\href{\doibase#1}}%
\providecommand \bibAnnote [3]{%
  \BibitemShut{#1}%
  \begin{quotation}\noindent
    \textsc{Key:}\ #2\\\textsc{Annotation:}\ #3%
  \end{quotation}%
}%
\providecommand \bibAnnoteFile [2]{%
  \IfFileExists{#2}{\bibAnnote {#1} {#2} {\input{#2}}}{}%
}%
\providecommand \typeout [0]{\immediate \write \m@ne }%
\providecommand \selectlanguage [0]{\@gobble}%
\providecommand \bibinfo [0]{\@secondoftwo}%
\providecommand \bibfield [0]{\@secondoftwo}%
\providecommand \translation [1]{[#1]}%
\providecommand \BibitemOpen[0]{}%
\providecommand \bibitemStop [0]{}%
\providecommand \bibitemNoStop [0]{.\EOS\space}%
\providecommand \EOS [0]{\spacefactor3000\relax}%
\providecommand \BibitemShut [1]{\csname bibitem#1\endcsname}%
\bibitem{Berger:1996}%
  \BibitemOpen
  \bibfield{author}{%
  \bibinfo {author} {\bibfnamefont{L.}~\bibnamefont{Berger}},\ }%
  \bibfield{journal}{%
  \Doi{10.1103/PhysRevB.54.9353}{\bibinfo {journal} {Phys. Rev. B}}\ }%
  \textbf{\bibinfo {volume} {54}},\ \bibinfo {pages} {9353} (\bibinfo {year} {1996})%
 \bibitem{SLonczewski:1989}%
  \BibitemOpen
  \bibfield{author}{%
  \bibinfo {author} {\bibfnamefont{J.~C.}\ \bibnamefont{Slonczewski}},\ }%
  \bibfield{journal}{%
  \Doi{10.1103/PhysRevB.39.6995}{\bibinfo {journal} {Phys. Rev. B}}\ }%
  \textbf{\bibinfo {volume} {39}},\ \bibinfo {pages} {6995} (\bibinfo {year} {1989})
  \bibAnnoteFile{NoStop}{SLonczewski:1989}%
\bibitem{Slonczewski:1996}%
  \BibitemOpen
  \bibfield{author}{%
  \bibinfo {author} {\bibfnamefont{J.}~\bibnamefont{Slonczewski}},\ }%
  \bibfield{journal}{%
  \bibinfo {journal} {J. Mag. Mag. Mater.}\ }%
  \textbf{\bibinfo {volume} {159}},\ \bibinfo {pages} {L1} (\bibinfo {year}
  {1996})%
  \bibAnnoteFile{NoStop}{Slonczewski:1996}%
\bibitem{Kaka:2005}%
  \BibitemOpen
  \bibfield{author}{%
  \bibinfo {author} {\bibfnamefont{S.}~\bibnamefont{Kaka}}, \bibinfo {author}
  {\bibfnamefont{M.~R.}\ \bibnamefont{Pufall}}, \bibinfo {author}
  {\bibfnamefont{W.~H.}\ \bibnamefont{Rippard}}, \bibinfo {author}
  {\bibfnamefont{T.~J.}\ \bibnamefont{Silva}}, \bibinfo {author}
  {\bibfnamefont{S.~E.}\ \bibnamefont{Russek}},\ and\ \bibinfo {author}
  {\bibfnamefont{J.~A.}\ \bibnamefont{Katine}},\ }%
  \bibfield{journal}{%
  \Doi{10.1038/nature04035}{\bibinfo {journal} {Nature}}\ }%
  \textbf{\bibinfo {volume} {437}},\ \bibinfo {pages} {389} (\bibinfo {year}
  {2005})%
  \bibAnnoteFile{NoStop}{Kaka:2005}%
\bibitem{Slavin:2009}%
  \BibitemOpen
  \bibfield{author}{%
  \bibinfo {author} {\bibfnamefont{A.}~\bibnamefont{Slavin}}\ and\ \bibinfo
  {author} {\bibfnamefont{V.}~\bibnamefont{Tiberkevich}},\ }%
  \bibfield{journal}{%
  \bibinfo {journal} {IEEE Trans. Magn.}\ }%
  \textbf{\bibinfo {volume} {45}},\ \bibinfo {pages} {1875} (\bibinfo {year} {2009})%
 \bibitem{Tsoi:2007}%
  \BibitemOpen
  \bibfield{author}{%
  \bibinfo {author} {\bibfnamefont{Z.}~\bibnamefont{Wei}}, \bibinfo {author}
  {\bibfnamefont{A.}~\bibnamefont{Sharma}}, \bibinfo {author}
  {\bibfnamefont{A.~S.}\ \bibnamefont{Nunez}}, \bibinfo {author}
  {\bibfnamefont{P.~M.}\ \bibnamefont{Haney}}, \bibinfo {author}
  {\bibfnamefont{R.~A.}\ \bibnamefont{Duine}}, \bibinfo {author}
  {\bibfnamefont{J.}~\bibnamefont{Bass}}, \bibinfo {author}
  {\bibfnamefont{A.~H.}\ \bibnamefont{MacDonald}},\ and\ \bibinfo {author}
  {\bibfnamefont{M.}~\bibnamefont{Tsoi}},\ }%
  \bibfield{journal}{%
  \bibinfo {journal} {Phys. Rev. Lett.}\ }%
  \textbf{\bibinfo {volume} {98}},\ \bibinfo {pages} {116603} (\bibinfo {year}
  {2007})%
  \bibAnnoteFile{NoStop}{Tsoi:2007}%
\bibitem{Urazhdin:2007}%
  \BibitemOpen
  \bibfield{author}{%
  \bibinfo {author} {\bibfnamefont{S.}~\bibnamefont{Urazhdin}}\ and\ \bibinfo
  {author} {\bibfnamefont{N.}~\bibnamefont{Anthony}},\ }%
  \bibfield{journal}{%
  \bibinfo {journal} {Phys. Rev. Lett.}\ }%
  \textbf{\bibinfo {volume} {99}},\ \bibinfo {pages} {046602} (\bibinfo {year}
  {2007})%
\bibitem{tang:122504}%
  \BibitemOpen
  \bibfield{author}{%
  \bibinfo {author} {\bibfnamefont{X.-L.}\ \bibnamefont{Tang}}, \bibinfo
  {author} {\bibfnamefont{H.-W.}\ \bibnamefont{Zhang}}, \bibinfo {author}
  {\bibfnamefont{H.}~\bibnamefont{Su}}, \bibinfo {author}
  {\bibfnamefont{Z.-Y.}\ \bibnamefont{Zhong}},\ and\ \bibinfo {author}
  {\bibfnamefont{Y.-L.}\ \bibnamefont{Jing}},\ }%
  \bibfield{journal}{%
  \Doi{10.1063/1.2786592}{\bibinfo {journal} {Appl. Phys. Lett.}}\ }%
  \textbf{\bibinfo {volume} {91}},\ \bibinfo {eid} {122504} (\bibinfo {year}
  {2007})
\bibitem{Dai08}%
  \BibitemOpen
  \bibfield{author}{%
  \bibinfo {author} {\bibfnamefont{N.~V.}\ \bibnamefont{Dai}}, \bibinfo
  {author} {\bibfnamefont{N.~C.}\ \bibnamefont{Thuan}}, \bibinfo {author}
  {\bibfnamefont{L.~V.}\ \bibnamefont{Hong}}, \bibinfo {author}
  {\bibfnamefont{N.~X.}\ \bibnamefont{Phuc}}, \bibinfo {author}
  {\bibfnamefont{Y.~P.}\ \bibnamefont{Lee}}, \bibinfo {author}
  {\bibfnamefont{S.~A.}\ \bibnamefont{Wolf}},\ and\ \bibinfo {author}
  {\bibfnamefont{D.~N.~H.}\ \bibnamefont{Nam}},\ }%
  \bibfield{journal}{%
  \Doi{10.1103/PhysRevB.77.132406}{\bibinfo {journal} {Phys. Rev. B}}\ }%
  \textbf{\bibinfo {volume} {77}},\ \bibinfo {eid} {132406} (\bibinfo {year}
  {2008})
  \bibAnnoteFile{NoStop}{Dai08}%
\bibitem{Tsoi-2008}%
  \BibitemOpen
  \bibfield{author}{%
  \bibinfo {author} {\bibfnamefont{J.}~\bibnamefont{Bass}}, \bibinfo {author}
  {\bibfnamefont{A.}~\bibnamefont{Sharma}}, \bibinfo {author}
  {\bibfnamefont{Z.}~\bibnamefont{Wei}},\ and\ \bibinfo {author}
  {\bibfnamefont{M.}~\bibnamefont{Tsoi}},\ }%
  \bibfield{journal}{%
  \bibinfo {journal} {Jour. of Magnetics (Korean Magnetics Society)}\ }%
  \textbf{\bibinfo {volume} {13}},\ \bibinfo {pages} {1} (\bibinfo {year}
  {2008})
\bibitem{Tsoi:1998}%
  \BibitemOpen
  \bibfield{author}{%
  \bibinfo {author} {\bibfnamefont{M.}~\bibnamefont{Tsoi}}, \bibinfo {author}
  {\bibfnamefont{A.~G.~M.}\ \bibnamefont{Jansen}}, \bibinfo {author}
  {\bibfnamefont{J.}~\bibnamefont{Bass}}, \bibinfo {author}
  {\bibfnamefont{W.-C.}\ \bibnamefont{Chiang}}, \bibinfo {author}
  {\bibfnamefont{M.}~\bibnamefont{Seck}}, \bibinfo {author}
  {\bibfnamefont{V.}~\bibnamefont{Tsoi}},\ and\ \bibinfo {author}
  {\bibfnamefont{P.}~\bibnamefont{Wyder}},\ }%
  \bibfield{journal}{%
  \Doi{10.1103/PhysRevLett.80.4281}{\bibinfo {journal} {Phys. Rev. Lett.}}\ }%
  \textbf{\bibinfo {volume} {80}},\ \bibinfo {pages} {4281} (\bibinfo {year} {1998})%
  \bibAnnoteFile{NoStop}{Tsoi:1998}%
\bibitem{Grollier:2003}%
  \BibitemOpen
  \bibfield{author}{%
  \bibinfo {author} {\bibfnamefont{J.}~\bibnamefont{Grollier}}, \bibinfo
  {author} {\bibfnamefont{V.}~\bibnamefont{Cros}}, \bibinfo {author}
  {\bibfnamefont{H.}~\bibnamefont{Jaffr\`es}}, \bibinfo {author}
  {\bibfnamefont{A.}~\bibnamefont{Hamzic}}, \bibinfo {author}
  {\bibfnamefont{J.~M.}\ \bibnamefont{George}}, \bibinfo {author}
  {\bibfnamefont{G.}~\bibnamefont{Faini}}, \bibinfo {author}
  {\bibfnamefont{J.}~\bibnamefont{Ben~Youssef}}, \bibinfo {author}
  {\bibfnamefont{H.}~\bibnamefont{Le~Gall}},\ and\ \bibinfo {author}
  {\bibfnamefont{A.}~\bibnamefont{Fert}},\ }%
  \bibfield{journal}{%
  \Doi{10.1103/PhysRevB.67.174402}{\bibinfo {journal} {Phys. Rev. B}}\ }%
  \textbf{\bibinfo {volume} {67}},\ \bibinfo {pages} {174402} (\bibinfo {year} {2003})%
\bibitem{stiles-2008-320}%
  \BibitemOpen
  \bibfield{author}{%
  \bibinfo {author} {\bibfnamefont{D.~C.}\ \bibnamefont{Ralph}}\ and\ \bibinfo
  {author} {\bibfnamefont{M.~D.}\ \bibnamefont{Stiles}},\ }%
  \bibfield{journal}{%
  \bibinfo {journal} {J. Mag. Mag. Mater.}\ }%
  \textbf{\bibinfo {volume} {320}},\ \bibinfo {pages} {1190} (\bibinfo {year}
  {2008})
\bibitem{Basset08}%
  \BibitemOpen
  \bibfield{author}{%
  \bibinfo {author} {\bibfnamefont{J.}~\bibnamefont{Basset}}, \bibinfo {author}
  {\bibfnamefont{A.}~\bibnamefont{Sharma}}, \bibinfo {author}
  {\bibfnamefont{Z.}~\bibnamefont{Wei}}, \bibinfo {author}
  {\bibfnamefont{J.}~\bibnamefont{Bass}},\ and\ \bibinfo {author}
  {\bibfnamefont{M.}~\bibnamefont{Tsoi}},\ }%
  \bibfield{journal}{%
  \Doi{10.1117/12.798220}{\bibinfo {journal} {Proc. of SPIE, Spintronics}}\ }%
  \textbf{\bibinfo {volume} {7036}},\ \bibinfo {eid} {703605} (\bibinfo {year}
  {2008})
  \bibAnnoteFile{NoStop}{Basset08}%
\bibitem{tserkovnyak-2006}%
  \BibitemOpen
  \bibfield{author}{%
  \bibinfo {author} {\bibfnamefont{Y.}~\bibnamefont{Tserkovnyak}}, \bibinfo
  {author} {\bibfnamefont{H.~J.}\ \bibnamefont{Skadsem}}, \bibinfo {author}
  {\bibfnamefont{A.}~\bibnamefont{Brataas}},\ and\ \bibinfo {author}
  {\bibfnamefont{G.~E.~W.}\ \bibnamefont{Bauer}},\ }%
  \bibfield{journal}{%
  \bibinfo {journal} {Phys. Rev. B}\ }%
  \textbf{\bibinfo {volume} {74}},\ \bibinfo {pages} {144405} (\bibinfo {year}
  {2006})
\bibitem{Nunez:2005}%
  \BibitemOpen
  \bibfield{author}{%
  \bibinfo {author} {\bibfnamefont{A.}~\bibnamefont{Nunez}}, \bibinfo {author}
  {\bibfnamefont{R.}~\bibnamefont{Duine}}, \bibinfo {author}
  {\bibfnamefont{P.}~\bibnamefont{Haney}},\ and\ \bibinfo {author}
  {\bibfnamefont{A.}~\bibnamefont{MacDonald}},\ }%
  \bibfield{journal}{%
  \bibinfo {journal} {Phys. Rev. B}\ }%
  \textbf{\bibinfo {volume} {73}},\ \bibinfo {pages} {214426} (\bibinfo {year}
  {2006})%
  \bibAnnoteFile{NoStop}{Nunez:2005}%
\bibitem{Haney:2007(2)}%
  \BibitemOpen
  \bibfield{author}{%
  \bibinfo {author} {\bibfnamefont{P.~M.}\ \bibnamefont{Haney}}\ and\ \bibinfo
  {author} {\bibfnamefont{A.~H.}\ \bibnamefont{MacDonald}},\ }%
  \bibfield{journal}{%
  \Doi{10.1103/PhysRevLett.100.196801}{\bibinfo {journal} {Phys. Rev.
  Lett.}}\ }%
  \textbf{\bibinfo {volume} {100}},\ \bibinfo {eid} {196801} (\bibinfo {year}
  {2008})
  \bibAnnoteFile{NoStop}{Haney:2007(2)}%
\bibitem{xu:226602}%
  \BibitemOpen
  \bibfield{author}{%
  \bibinfo {author} {\bibfnamefont{Y.}~\bibnamefont{Xu}}, \bibinfo {author}
  {\bibfnamefont{S.}~\bibnamefont{Wang}},\ and\ \bibinfo {author}
  {\bibfnamefont{K.}~\bibnamefont{Xia}},\ }%
  \bibfield{journal}{%
  \Doi{10.1103/PhysRevLett.100.226602}{\bibinfo {journal} {Phys. Rev.
  Lett.}}\ }%
  \textbf{\bibinfo {volume} {100}},\ \bibinfo {eid} {226602} (\bibinfo {year}
  {2008})
\bibitem{gomo:2008E}%
  \BibitemOpen
  \bibfield{author}{%
  \bibinfo {author} {\bibfnamefont{E.~V.}\ \bibnamefont{Gomonay}}\ and\
  \bibinfo {author} {\bibfnamefont{V.~M.}\ \bibnamefont{Loktev}},\ }%
  \bibfield{journal}{%
  \Doi{10.1063/1.2889408}{\bibinfo {journal} {Sov. J. Low Temp. Phys.}}\ }%
  \textbf{\bibinfo {volume} {34}},\ \bibinfo {pages} {198} (\bibinfo {year}
  {2008})
  \bibAnnoteFile{NoStop}{gomo:2008E}%
\bibitem{Haney:2007}%
  \BibitemOpen
  \bibfield{author}{%
  \bibinfo {author} {\bibfnamefont{P.~M.}\ \bibnamefont{Haney}}, \bibinfo
  {author} {\bibfnamefont{D.}~\bibnamefont{Waldron}}, \bibinfo {author}
  {\bibfnamefont{R.~A.}\ \bibnamefont{Duine}}, \bibinfo {author}
  {\bibfnamefont{A.~S.~N.}\ \bibnamefont{{n}ez}}, \bibinfo {author}
  {\bibfnamefont{H.}~\bibnamefont{Guo}},\ and\ \bibinfo {author}
  {\bibfnamefont{A.~H.}\ \bibnamefont{MacDonald}},\ }%
  \bibfield{journal}{%
  \Doi{10.1103/PhysRevB.75.174428}{\bibinfo {journal} {Phys. Rev. B}}\ }%
  \textbf{\bibinfo {volume} {75}},\ \bibinfo {eid} {174428} (\bibinfo {year}
  {2007})
  \bibAnnoteFile{NoStop}{Haney:2007}%
\bibitem{haney:2007(3)}%
  \BibitemOpen
  \bibfield{author}{%
  \bibinfo {author} {\bibfnamefont{P.~M.}\ \bibnamefont{Haney}}, \bibinfo
  {author} {\bibfnamefont{D.}~\bibnamefont{Waldron}}, \bibinfo {author}
  {\bibfnamefont{R.~A.}\ \bibnamefont{Duine}}, \bibinfo {author}
  {\bibfnamefont{A.~S.~N.}\ \bibnamefont{{n}ez}}, \bibinfo {author}
  {\bibfnamefont{H.}~\bibnamefont{Guo}},\ and\ \bibinfo {author}
  {\bibfnamefont{A.~H.}\ \bibnamefont{MacDonald}},\ }%
  \bibfield{journal}{%
  \Doi{10.1103/PhysRevB.76.024404}{\bibinfo {journal} {Phys. Rev. B}}\ }%
  \textbf{\bibinfo {volume} {76}},\ \bibinfo {eid} {024404} (\bibinfo {year}
  {2007})
\bibitem{Bar'yakhtar:1984E}%
  \BibitemOpen
  \bibfield{author}{%
  \bibinfo {author} {\bibfnamefont{V.~G.}\ \bibnamefont{Bar'yakhtar}},\ }%
  \bibfield{journal}{%
  \bibinfo {journal} {Sov. Phys. --- JETP}\ }%
  \textbf{\bibinfo {volume} {60}},\ \bibinfo {pages} {863} (\bibinfo {year}
  {1984})%
\bibitem{Bar:1968E}%
  \BibitemOpen
  \bibfield{author}{%
  \bibinfo {author} {\bibfnamefont{A.~I.}\ \bibnamefont{Akhiezer}}, \bibinfo
  {author} {\bibfnamefont{V.~G.}\ \bibnamefont{Bar'yakhtar}},\ and\ \bibinfo
  {author} {\bibfnamefont{S.~V.}\ \bibnamefont{Peletminskii}},\ }%
  \emph{\bibinfo {title} {Spin Waves}},\ \bibinfo {edition} {Interscience
  (Wiley)}\ ed.,\ \bibinfo {series} {North-Holland Series in Low Temperature
  Physics}, Vol.~\bibinfo {volume} {1}\ (\bibinfo {publisher} {North-Holland},\
  \bibinfo {address} {Amsterdam},\ \bibinfo {year} {1968})%
  \bibAnnoteFile{NoStop}{Bar:1968E}%
\bibitem{Bar-june:1979E}%
  \BibitemOpen
  \bibfield{author}{%
  \bibinfo {author} {\bibfnamefont{I.}~\bibnamefont{Bar’yakhtar}}\ and\
  \bibinfo {author} {\bibfnamefont{B.}~\bibnamefont{Ivanov}},\ }%
  \bibfield{journal}{%
  \bibinfo {journal} {Sov. --- J. Low Temp. Phys.}\ }%
  \textbf{\bibinfo {volume} {5}},\ \bibinfo {pages} {361} (\bibinfo {year}
  {1979})%
  \bibAnnoteFile{NoStop}{Bar-june:1979E}%
\bibitem{Note1}%
  \BibitemOpen
  \bibinfo {note} {~In the bilayer structure the orientation of AFM vector is
  governed by competition between the intrinsic magnetic anisotropy (\ref
  {anisotropy}) and exchange coupling with the adjacent FM layer. However, in
  the typical cases an effective field of exchange coupling is much smaller
  than the spin-flop field of AFM and thus can be neglected.}%
  \bibAnnoteFile{Stop}{Note1}%
\bibitem{Kiselev:2003}%
  \BibitemOpen
  \bibfield{author}{%
  \bibinfo {author} {\bibfnamefont{S.~I.}\ \bibnamefont{Kiselev}}, \bibinfo
  {author} {\bibfnamefont{J.~C.}\ \bibnamefont{Sankey}}, \bibinfo {author}
  {\bibfnamefont{I.~N.}\ \bibnamefont{Krivorotov}}, \bibinfo {author}
  {\bibfnamefont{N.~C.}\ \bibnamefont{Emley}}, \bibinfo {author}
  {\bibfnamefont{R.~J.}\ \bibnamefont{Schoelkopf}}, \bibinfo {author}
  {\bibfnamefont{R.~A.}\ \bibnamefont{Buhrman}},\ and\ \bibinfo {author}
  {\bibfnamefont{D.~C.}\ \bibnamefont{Ralph}},\ }%
  \bibfield{journal}{%
  \Doi{10.1038/nature01967}{\bibinfo {journal} {Nature}}\ }%
  \textbf{\bibinfo {volume} {425}},\ \bibinfo {pages} {380} (\bibinfo {year}
  {2003}).%
\bibitem{Gulyaev:2005}%
  \BibitemOpen
  \bibfield{author}{%
  \bibinfo {author} {\bibfnamefont{Y.~V.}\ \bibnamefont{Gulyaev}}, \bibinfo
  {author} {\bibfnamefont{P.~E.}\ \bibnamefont{Zil’berman}}, \bibinfo {author}
  {\bibfnamefont{E.~M.}\ \bibnamefont{Epshtein}},\ and\ \bibinfo {author}
  {\bibfnamefont{R.~J.}\ \bibnamefont{Elliott}},\ }%
  \bibfield{journal}{%
  \bibinfo {journal} {Sov. Phys.--- JETP}\ }%
  \textbf{\bibinfo {volume} {100}},\ \bibinfo {pages} {1005} (\bibinfo {year}
  {2005}).%
\bibitem{Bogoluybov:1998}%
  \BibitemOpen
  \bibfield{author}{%
  \bibinfo {author} {\bibfnamefont{N.~N.}\ \bibnamefont{Bogolyubov}}\ and\
  \bibinfo {author} {\bibfnamefont{Y.~A.}\ \bibnamefont{Mitropolskii}},\ }%
  \emph{\bibinfo {title} {Asymptotic methods in the theory of nonlinear
  oscillations}}\ (\bibinfo {publisher} {Taylor and Francis, Inc.},\ \bibinfo
  {year} {1961})\ \bibinfo {note} {548 p.}%
  \bibAnnoteFile{Stop}{Bogoluybov:1998}%
\bibitem{Endoh:1973}%
  \BibitemOpen
  \bibfield{author}{%
  \bibinfo {author} {\bibfnamefont{Y.}~\bibnamefont{Endoh}}, \bibinfo {author}
  {\bibfnamefont{G.}~\bibnamefont{Shirane}}, \bibinfo {author}
  {\bibfnamefont{Y.}~\bibnamefont{Ishikawa}},\ and\ \bibinfo {author}
  {\bibfnamefont{K.}~\bibnamefont{Tajima}},\ }%
  \bibfield{journal}{%
  \bibinfo {journal} {Solid State Commun.}\ }%
  \textbf{\bibinfo {volume} {13}},\ \bibinfo {pages} {1179} (\bibinfo {year}
  {1973})%
  \bibAnnoteFile{NoStop}{Endoh:1973}%
\bibitem{Ivanov:2009}%
  \BibitemOpen
  \bibfield{author}{%
  \bibinfo {author} {\bibfnamefont{A.~V.}\ \bibnamefont{Kimel}}, \bibinfo
  {author} {\bibfnamefont{B.~A.}\ \bibnamefont{Ivanov}}, \bibinfo {author}
  {\bibfnamefont{R.~V.}\ \bibnamefont{Pisarev}}, \bibinfo {author}
  {\bibfnamefont{P.~A.}\ \bibnamefont{Usachev}}, \bibinfo {author}
  {\bibfnamefont{A.}~\bibnamefont{Kirilyuk}},\ and\ \bibinfo {author}
  {\bibfnamefont{T.}~\bibnamefont{Rasing}},\ }%
  \bibfield{journal}{%
  \Doi{doi:10.1038/nphys1369}{\bibinfo {journal} {Nature Physics}}\ }%
  \textbf{\bibinfo {volume} {5}},\ \bibinfo {pages} {727} (\bibinfo {year}
  {2009})%
\end{thebibliography}

%
\end{document}